\documentclass[12pt,preprint]{aastex}
\slugcomment{Submitted to ApJ-v3: ~~\today}
\begin{document}

\title{SIMULTANEOUS CONSTRAINTS ON THE SPECTRUM OF THE EXTRAGALACTIC BACKGROUND
LIGHT AND THE INTRINSIC TeV SPECTRA OF Mrk 421, Mrk 501, and H1426+428}


\author{Eli Dwek}
\affil{Laboratory for Astronomy and Solar Physics \\ NASA Goddard Space Flight
Center \\
Greenbelt, MD 20771 \\  e-mail: eli.dwek@gsfc.nasa.gov}

\and

\author{Frank Krennrich}
\affil{Department of Physics \& Astronomy \\ Iowa State University \\  Ames, IA,
50011, USA}

\begin{abstract}
Very high energy ($\sim$ TeV) $\gamma$-rays from blazars are attenuated by photons 
from the extragalactic background light (EBL). Observations of blazars can
therefore 
provide an ideal opportunity for determining the EBL 
intensity if their intrinsic spectrum is known. Conversely,  knowledge of the EBL 
intensity can be used to determine the intrinsic blazar spectrum. Unfortunately, 
neither the EBL intensity nor the intrinsic blazar spectrum is known with high 
enough precision to accurately derive one quantity from the other. 
In this paper we use the most recent data on the EBL to construct twelve
different realizations representing all possible permutations between EBL limits
and the detections in the different wavelength regions. These realizations
explore a significantly larger range of allowable EBL spectra than any previous
studies. We show that these realizations can be used to explore the effects of
the EBL on the inferred spectra of blazars. Concentrating on the 
two relatively nearby ($z \approx$ 0.03) blazars Mrk~421 and 501, we derive
their intrinsic spectra and peak $\gamma$-ray energies for the different EBL
realizations. Some EBL spectra give rise to ``unphysical" intrinsic blazar
spectra, characterized by an exponential rise at high TeV energies. We use the
F-test to show that some of these exponential rises are statistically
insignificant. Consequently, statements regarding the existence of a "IR
background-TeV $\gamma$-ray crisis" are unfounded on the basis of our present
knowledge of the EBL. EBL spectra that do give rise to unphysical blazar spectra
are regarded as invalid realizations of the EBL. Those that do not, define new
constraints on 
the EBL spectrum, and are used to derive new limits on the intensity and the peak 
$\gamma$-ray energy of these  two blazars. In particular, we derive an upper
limit of $\sim$ 15~nW~m$^{-2}$~sr$^{-1}$ on the 60~$\mu$m EBL intensity, and
find the peak energies of the two Mrk~421 and Mrk~501 sources to be very
similar, between 0.5--1.2~TeV and 0.8--2.5~TeV, respectively. We also show that
the intrinsic spectrum of Mrk~421 during a period of intense flaring activity
has a peak energy that seems to shift to higher energies at higher flux states.
Finally, we show that most EBL realizations show that the spectrum of the most
distant blazar H1426+428 ($z$ = 0.129) peaks between $\sim$ 1 and 5~TeV, with
some suggesting peaks below 400~GeV and above 10~TeV. These results provide
important constraints on the different particle acceleration mechanisms and the
generation of the $\gamma$-ray emission. Finally we also show that uncertainties
in the absolute calibration of the $\gamma$-ray energies with atmospheric
Cherenkov telescopes have an important impact on the intrinsic blazar spectra.
These systematic uncertainties will be improved with the next generations of
telescopes which will also cover a wider range of $\gamma$-ray energies,
providing further insights into the intrinsic spectrum of TeV blazars.
\end{abstract}

\section{INTRODUCTION}
Observations of blazars  at TeV $\gamma$-ray energies provide a view of the high
energy 
tail of blazar spectra, and are instrumental for probing non-thermal phenomena
in  jets 
of active galactic nuclei. 
These spectra provide important constraints on the particle acceleration
mechanism that 
generates $\gamma$-ray photons in the jets of active galactic nuclei, the
cooling processes 
by which particles dissipate their energy into electromagnetic radiation, and on
the 
astrophysical conditions in the acceleration region. TeV observations therefore
might offer a 
powerful way for discriminating between different $\gamma$-ray emission models,
recently reviewed by Coppi (2003). 

To the dismay of $\gamma$-ray astronomers, high energy $\gamma$-rays traveling
cosmological 
distances are attenuated en route to Earth by  $\gamma+\gamma \rightarrow e^+ +
e^-$ interactions 
with photons from the extragalactic background light (EBL; Gould \& Schr\'eder 
1967; Stecker, de Jager, \& Salamon 1992).  Consequently,  particle acceleration
mechanisms and $\gamma$-ray 
production models can only be tested after the observed blazar is corrected for
this attenuation.

However, to the delight of infrared (IR) astronomers, the attenuation of TeV
$\gamma$-rays can be used 
to constrain the intensity and spectrum of the EBL in the wavelength region
where its direct 
measurement is most difficult.
Defined here as the diffuse background light in the 0.1 to 1000~$\mu$m wavelength
regime, the EBL 
is the second most important radiative energy density in the universe after the
cosmic 
microwave background. It is the repository of all radiative energy inputs in the
universe 
since the epoch of decoupling (Partridge \& Peebles 1967; Hauser \& Dwek 2001).
It therefore contains important information on the release of nuclear,
gravitational, and possible exotic forms of energy throughout most of the
history of the 
universe. Unfortunately, direct measurement of the EBL in the $\sim$ 5 to
60~$\mu$m wavelength region is greatly hampered by foreground emission from
interplanetary dust (Hauser et al. 1998; 
Kelsall et al. 1998). Consequently, TeV $\gamma-$ray observations may be the
only way to 
constrain or determine the EBL in this wavelength regime.

Early studies (e.g. Stecker, de Jager, \& Salamon 1992; Dwek \& Slavin 1994;
Biller et al. 1995) assumed 
that the intrinsic spectrum of blazars is a power law in energy $E$ of the form 
$dN/dE \propto E^{-\alpha}$, and searched for a ``break" in the spectrum that
could be attributed 
to attenuation by the EBL. In a more recent paper, 
Costamante, Aharonian, Horns, \& Ghisellini (2004) used assumptions that reach
beyond these early 
attempts relying on a "break" in the spectrum: they require that the spectrum be
"smooth", to derive limits on the EBL. 
 Some of the earlier  approaches have been largely discarded  with the
realization that blazar spectra are 
considerably more complex, characterized by an energy  dependent spectral index
and variable intensity 
(Samuelson et al. 1998; Aharonian et al. 1999,  2002a; Djannati-Ata\"{\i} et al.
1999; Krennrich et al. 1999, 2001, 2002). Construction of the intrinsic blazar spectrum requires therefore a detailed knowledge of the $\gamma$-ray production mechanism. For example, Coppi  \&
Aharonian (1999) suggested using contemporaneous X-ray/$\gamma$-ray variations in the blazar emission in conjuncture with a synchrotron self-Compton emission model to estimate the intrinsic TeV spectrum of the blazar from its X-ray spectrum. Differences between this intrinsic spectrum and the observed can then be used to estimate the amount of intergalactic absorption and the intensity of the EBL.
However, there still is no consensus on the $\gamma$-ray emission model, which prompted an alternate  approach of using either theoretical EBL models, or
observed EBL limits and detections in order 
to place constraints on the  intrinsic blazar spectra (Guy et al. 2000;  Renault
et al. 2001; 
Dwek \& de Jager 2001;  Primack, Somerville, Bullock, \& Devriendt 2001; Kneiske, Mannheim, \& Hartmann 2002; Kneiske, Bretz, Mannheim, \& Hartmann 2004; and Wright 2004 ).

In this paper we use the most recent data on the EBL to construct twelve
different realizations, representing all possible permutations between EBL
limits and the detections in the different wavelength regions, to explore the
effects of the EBL on the inferred spectra of blazars.
We focus on the determination of the intrinsic energy spectra of the two
most prominent and nearby TeV blazars, Mrk~421 and  Mrk~501  which have been
measured with good
statistical precision at energies  between 260~GeV and 20~TeV (Samuelson et al.
1998; 
Aharonian et al. 1999, 2002a; Krennrich et al. 1999, 2001, 2002;
Djannati-Ata\"{\i} et al. 1999). Some claimed EBL detections, notably the 60 $\mu$m intensity derived by Finkbeiner, Davis, \& Schlegel (2000),  produce a pile-up of photons in the highest energy bins of the intrinsic blazar spectrum. This pile-up has led to claims of the existence of a ``TeV--IR
Background crisis" 
(Protheroe \& Meyer 2000). However, the EBL spectrum required for calculating the $\gamma$-ray opacity towards these
sources,  
consist largely  of  upper and lower limits and few  3$\rm \sigma$  level
detections in the 0.1~-~200~$\mu$m 
wavelength region (Hauser \& Dwek 2001). Uncertainties in the EBL spectrum
propagate exponentially into 
the derived blazar spectrum, a fact that has been largely ignored by previous
investigators (e.g. Konopelko 
et al. 2003). Consequently, any statements claiming the existence of a ``TeV--IR
Background crisis" are unfounded considering the uncertainties in the EBL intensity and the observed $\gamma$-ray fluxes in the highest energy bins. 



In spite of the uncertainties in the detailed spectral behavior of the EBL, its
general double--peak nature is well established. 
The peaks at $\sim$ 1 and $\sim$ 100~$\mu$m, are associated with energy releases
from stars 
and dust, respectively. The drop in the EBL spectrum between these two peaks 
($\lambda \approx\ 5-60\ \mu$m) moderates the rise in the TeV opacity with
$\gamma-$ray energies, 
creating an opportunity for extracting the intrinsic 200 GeV to 10 TeV spectrum
of these nearby blazars.

Using the observed detections and limits on the EBL we create a family of EBL
spectra representing 
different realizations of the EBL, bounding extreme limits in the UV optical and
IR-submillimeter 
wavelength regions. The EBL spectra used in our analysis cover a significantly
wider range of intensities than the two "extreme" spectra considered by de
Jager, and Stecker (2002),
Konopelko et al. (2003), and the range of EBL spectra recently derived by Kneiske et al. (2004). The family of EBL spectral templates are used to derive
the $\gamma$-ray optical 
depth to the observed blazars, to construct the intrinsic spectra of Mrk~421,
Mrk~501, and H1426+428, and
to examine spectral trends in the intrinsic variability of Mrk~421. We show that
several intrinsic 
spectra display "unphysical" characteristics, manifested as an exponential rise
with energy, that 
can be directly related to the spectrum of 
the EBL used for calculating the $\gamma$-ray attenuation.
Aharonian, Timokhin, \& Plyasheshnikov (2002b, ATP02) suggested that a pile-up of photons in the highest energy bins actually represents the  intrinsic blazar spectrum, and presented a model in which the $\sim$10-20~TeV photons are produced by  the comptonization of ambient optical photons by an ultrarelativistic jet. However, the exponential rises produced by some EBL realizations are unbounded, extending well beyond the highest energy bins in the blazar spectrum, and are therefore unlikely to be generated by the proposed ultrarelativistic jet-model.
 
The exponential rises in the blazar spectrum derived here result from correcting the observed $\gamma$-ray fluxes in the highest energy bins for EBL absorption. Uncertainties in the observations may affect the ``reality" of some of these exponential rises. We therefore employ the statistical F-test to examine the significance of these
rises. 
Blazar spectra that are confirmed to be unphysical are then used to set 
new limits on the EBL. We derive the peak energies in the well-behaved intrinsic
blazar spectra, and 
discuss the relevance of these results for blazar unification models (Costamante
et al. 2001). 

Another source of uncertainty in the blazar spectrum is the absolute energy of the $\gamma$-ray photons. We therefore examine the effect of these uncertainties on the intrinsic blazar spectra and on the
ability to constrain the EBL. Finally, we also examine whether the TeV optical
depth towards Mrk~421 and 501, which are located at an almost identical redshift
of 0.03, may differ due to statistical fluctuations in the number of sources
along their line of sight.

The paper is organized as follows: In \S2 we present the observed $\gamma-$ray
spectrum of Mrk~421, Mrk~501, 
and H1426+428 and the spectral variability of Mrk~421. 
The current status of the detection and limits of the EBL are described in \S3,
where we also present the different spectral realizations of the EBL. In \S4 we
derive the  $\gamma$-ray opacity for all the EBL realizations. The attenuation
corrected spectra of Mrk 421 and Mrk~501 are presented in \S5, and EBL
realizations leading to unphysical intrinsic spectra are identified. The
constraints on the EBL derived from analysis of the intrinsic blazar spectrum
are discussed in \S6. Blazar spectra corrected for absorption by the viable EBL
realizations are presented in \S7. A brief summary of 
the paper is presented in \S8.   

 \section{GAMMA-RAY DATA OF MRK~421, MRK~501 AND H1426+428}

The $\gamma$-ray data used for this study were obtained by two experimental
groups, the VERITAS  
collaboration using a single 10~m diameter imaging atmospheric Cerenkov
telescope at the Whipple
observatory (Finley et al. 2001) in southern Arizona and the HEGRA collaboration 
using an array of smaller size  imaging atmospheric Cerenkov  telescopes 
(Daum et al. 1997) on La Palma, Canary Islands. 
The atmospheric Cerenkov technique is prone to uncertainties in the absolute
energy scale of the $\gamma$-ray photons resulting from  
systematic uncertainties in modeling the Earth's atmosphere, and uncertainties
in the temporal variations of its transparency. These uncertainties limit the
accuracy of the absolute energy 
calibration of the $\sim$~TeV  $\gamma$-rays to 15\%, and can have an important
effect on the correction applied to the observed blazar spectrum due to
absorption by the EBL. 
A consistency check of the magnitude of major systematic
uncertainties  of different atmospheric Cerenkov experiments is given by
measurements of 
the Crab Nebula, the standard candle of $\gamma$-ray astronomy. The energy
spectra of 
the Crab Nebula measured by  the Whipple telescope and the HEGRA detector are in
agreement within 
systematic errors (for further details see also Aharonian  et al. 2000; Hillas
et al. 1998). 

This study is primarily concerned with the two most prominent blazars in TeV
gamma-ray astronomy, Mrk~421 
and Mrk~501. Both received a lot of attention because of their episodes of
strong flaring activity. 
Furthermore, at redshifts of 0.031 and 0.034,
respectively, they are approximately at the same distance from the observer,
suggesting that absorption by the EBL should affect them similarly. 
 Mrk~501 was discovered as a  $\gamma$-ray blazar at TeV energies (Quinn et al.
1996) and to a 
great surprise underwent a strong and long-lasting flaring state during the
first half of 1997  
(Catanese et al. 1997;  Aharonian et al. 1997; Djannati-Ata\"{\i} et al. 1999; 
Protheroe et al. 1998).   The good statistics  of these data led to the
discovery of a first 
spectral feature in a TeV blazar, the detection of a cutoff manifested as a
strong deviation of d$N$/d$E$ from an extrapolated $E^{-2}$ power law, at
approximately 4~TeV 
(Samuelson et al. 1998).    Detailed spectral measurements by other groups
independently 
confirmed this cutoff feature (Aharonian et al. 1997, 1999; Djannati-Ata\"{\i}
et al. 1999).
The combined Whipple and HEGRA data cover an energy range between  260~GeV  and
 20~TeV,
making the energy spectrum of Mrk~501 an interesting constraint to the EBL density.

The second blazar  that we discuss in detail in this paper is Mrk~421,  which
was established as a TeV 
source by Punch et al. (1992).  Although strong and short  flares were observed
in 1995 and 
in 1996 (Gaidos et al. 1996), those data were not sufficient to unravel 
spectral features beyond a power law index (Zweerink et al. 1997; Krennrich et
al. 1999).     
This changed when an active episode of  long-lasting strong flares occurred for
 Mrk~421 in 2000/2001, 
providing the best statistics  ever recorded for a $\gamma$-ray blazar and also
showing a cutoff
at approximately 4~TeV (Krennrich et al. 2001;  Aharonian et al. 2002a). The
flaring data from the observing season of 2000/2001 have also been used to study
the flux spectral index as
a function of the flux level.  The data have been binned according to flux
levels resulting into 6 data sets at different flux levels (Krennrich et al.
2002).  These data have unraveled for the first time significant spectral
variability at TeV energies. 

The spectral data of Mrk~421 and Mrk~501 from the two experiments are
complementary.  
The Whipple 10~m telescope data start at 260~GeV extending out to 17~TeV for
Mrk~421 and 
 out to 12.1 TeV for Mrk~501,  whereas the HEGRA observations start at 560~GeV
and extend to 17~TeV for Mrk~421 and to 21~TeV for Mrk~501.   Due of the wealth
of statistics, the uncertainties in the data are dominated by systematic
uncertainties in the absolute calibration of the Whipple and HEGRA instruments.
Consequently, we refrain from combining the data and analyze them separately.

Finally, we present spectral data from a first more distant ($z$=0.129) TeV
blazar, H1426+428,
which was discovered as a TeV source by Horan et al. (2000, 2001a, 2001b, 2002)
and confirmed by 
Aharonian et al. (2002a) and Djannati-Ata\"{\i} et al. (2002).  We use the
spectral data 
from the HEGRA collaboration  (Aharonian et al. 2002a) and the Whipple
collaboration (Petry et al. 2002).  
Because of the limited  statistics of the spectra from both groups (5.8 and 5.5
sigma detections)
we combine the spectra (as was presented in Petry et al. 2002). This is possible
since
the statistical errors dominate over the systematic  uncertainties for this data
set.
Figure 1 summarizes the observed spectra of the three blazars, and presents the
observed spectral variability of Mrk~421.

\section{REALIZATIONS OF THE EXTRAGALACTIC \\ BACKGROUND LIGHT}

\subsection{Observational Limits and Detections of the EBL}
The EBL is viewed through strong sources of foreground emission consisting of
reflected sunlight and thermal emission from zodiacal dust particles, resolved
and diffuse Galactic starlight, diffuse emission from interstellar dust, and
emission from resolved Galactic H~II regions and nearby galaxies (LMC, SMC). The
relative contribution of these different foreground emission components varies
significantly over the 0.1 to 1000 $\mu$m wavelength range of the EBL.

Removal of the zodiacal dust emission from the observed sky maps posed the
greatest challenge for the detection of the EBL in the 1 to 1000~$\mu$m
wavelength region. Thermal emission from the zodiacal cloud dominates the
foreground in the $\sim$ 5--100 $\mu$m wavelength region with an intensity of
$\nu I_{\nu} \sim 4 \times 10^3$ nW m$^{-2}$ sr$^{-1}$ between $\sim$ 15 and 30
$\mu$m in the direction of the Lockman Hole. This intensity is about a factor of
$\sim 100 - 10^3$ larger than the average EBL intensity expected from
nucleosynthesis arguments, illustrating the difficulty of separating this
component from the EBL. The subtraction of the  zodiacal dust emission was
described in detail by Kelsall et al. (1998). The procedure modeled the
variation in the sky intensity caused by the Earth's motion through the IPD
cloud and the DIRBE scanning pattern. Any zodiacal dust model is therefore
insensitive to any isotropic emission component of the cloud. The uncertainties
in the intensity of this component were determined from the variance in its
value, obtained by modeling the primary dust cloud with different geometrical
configurations which produced about equally good fits to the observed variations
in the sky brightness. 
The Kelsall et al. (1998) zodiacal light model (KZL model hereafter) succeeded
in subtracting 98\% of the thermal emission from the zodi cloud. The residual 5
to 100~$\mu$m emission exhibited a strong peak at 15--30~$\mu$m. It had large
systematic uncertainties and was far from isotropic, preventing its
identification as a component of the EBL. 

Arguing that the $\sim$ 15 to 30 $\mu$m residual must be mostly of solar system
origin (otherwise the observed TeV blazars would not be detectable), Gorjian et
al. (2000) used a zodiacal dust model similar to the KZL model, but imposed a
``very strong no-zodi principle"  (Wright 2001), requiring that the 25 $\mu$m
residual after the removal of the zodiacal emission be zero at high Galactic
latitudes. We will hereafter refer to this zodiacal light model as the GWZ
model. The resulting contribution of the GWZ model to the foreground emission is
therefore larger than that of the KZL model, resulting in lower values for the
EBL intensity in the 1.25 to 5~$\mu$m wavelength region.  

Galactic starlight is an important contributor to the foreground emission at
near infrared wavelengths  ($\lambda \approx$ 1--5 $\mu$m), and the removal of
this component from the DIRBE skymaps was discussed in detail by Arendt et al.
(1998). The systematic uncertainties in the $\sim$ 1 -- 5 $\mu$m residuals were
dominated by uncertainties in the model used to subtract the emission from
unresolved stars. Since then significant efforts have been undertaken to improve
the removal of the Galactic stellar emission component, resulting in the
detection of the EBL at 1.25, 2.2, and 3.5 $\mu$m (Dwek \& Arendt 1998, Wright
\& Reese 2000, Wright 2001, Cambr\'esy et al. 2001, Arendt \& Dwek 2003). Larger
values for the EBL at these wavelengths were obtained by Cambr\'esy et al.
(2001) and Matsumoto et al. (2000), who used the KZL--model to subtract the
zodiacal foreground. Use of the GWZ--model, which is characterized by a larger
contribution of the zodiacal dust cloud to the foreground emission, will give
rise to a lower EBL intensity. The dependence of the near--IR EBL intensity  on
the model used for the subtraction of the zodiacal emission is discussed by
Arendt \& Dwek (2003) and illustrated in Figure 6 of their paper. All {\it
COBE}/DIRBE derived EBL intensities at between $\sim$ 1 and 2~$\mu$m are
significantly higher than those derived by Madau \& Pozzetti (2000) from galaxy
number counts. This should not be too surprising, since the integrated light
from galaxies provides only a strict lower limit on the EBL intensity, even when
the integrated light from the resolved galaxies seems to have converged.
Bernstein, Freedman, \& Madore (2002) pointed out that a significant fraction of
the flux from resolved galaxies can remain undetected, since the overlapping
wings of these galaxies can form a truly diffuse background which will be missed
in standard galaxy aperture photometry.  

At mid--IR wavelengths the EBL has a lower limit at 15 $\mu$m, derived from
galaxy number counts obtained with the {\it Infrared Space Observatory} ({\it
ISO}) satellite (Elbaz et al. 2002; Metcalfe et al. 2003). The integrated flux from 24~$\mu$m sources detected in deep surveys with the {\it Spitzer} satellite gives a lower limit of 1.9$\pm$0.6~nW m$^{-2}$ on the EBL intensity, which extrapolated to fainter flux densities provides an estimated EBL intensity of 2.7$^{+1.1}_{-0.7}$~nW~m$^{-2}$~sr$^{-1}$ at that wavelength (Papovich et al. 2004). An upper limit of about 5 nW m$^{-2}$
sr$^{-1}$ was derived by Stanev \& Franceschini (1998), Renault et al. (2001),
and Dwek \& de Jager (2001) from TeV $\gamma$-ray considerations.  At
far--infrared wavelengths, the EBL has been detected at $\sim$ 200 to 1000
$\mu$m  by Puget et al. (1996), Fixsen et al. (1998), at 140 and 240 $\mu$m by
Hauser et al. (1998), and at 100 $\mu$m by Lagache et al. (2000). Hauser et al.
(1998) reported the 140~$\mu$m EBL intensity derived using the DIRBE photometric
calibration. A somewhat lower value (but consistent  with the DIRBE calibration)
is derived if the FIRAS photometric scale is used in the calibration (Hauser et
al. 1998, Hauser \& Dwek 2001). 

At UV wavelengths Gardner et al. (2000) presented the integrated light obtained
from deep galaxy counts using the Space Telescope Imaging Spectrograph (STIS)
combined with counts obtained with the FOCA balloon-borne telescope (Milliard et
al. 1992). Bernstein, Freedman, \& Madore (2002) reported the detection of the
EBL at 0.3, 0.55, and 0.8~$\mu$m, at levels that are higher than the results of
Madau \& Pozzetti (2000), results that were disputed by Matilla (2003). Current
limits and detections of the EBL are presented in Table~1, and depicted in Figure 2.


\subsection{Template EBL Spectrum}

Given the general double--peak nature of the EBL and the uncertainties in its
spectral intensity we created a family of EBL spectra spanning extreme possible
combinations of relative peak values. 
Altogether we constructed twelve template spectra, representing different
realizations of the EBL, by fitting polynomials to all possible combinations of
the following spectral components: 
\noindent
\begin{enumerate}
\item Three stellar components consisting of: (1) {\it high--UV: }  defined by
the 0.1595 and 0.2365~$\mu$m data from Gardner et al. (2000), and the Cambr\'esy
et al. (2001) determinations that used the KZL zodi-light model; (2) {\it
mid--UV: }  defined by the 0.1595 and 0.2365  $\mu$m data from Gardner et al.
(2000), and the Wright (2001) EBL intensity derived by using the GWZ zodi-light
model; and (3)
{\it low--UV: }  defined by the 0.1595 and 0.2365  $\mu$m data from Gardner et
al. (2000), and the number counts of Madau \& Pozzetti (2000). 
The three realizations of the stellar components are driven by the near--IR and
the 0.1595 and 0.2365  $\mu$m constraints, and are compatible with the claimed
detections by Bernstein et al. (2002), considering the uncertainties in their values. 
\item Two mid--IR components, which were defined by the uncertainties in the
15~$\mu$m lower limit determined by Elbaz et al. (2002) from {\it ISO}
observations. The high mid--IR EBL is represented by the nominal 15~$\mu$m
intensity $+\ 3\sigma$, and the low mid--IR EBL is represented by the nominal
15~$\mu$m intensity $-\ 3\sigma$.  
 \item Two far--IR components, defined by the 2 different calibration of the
DIRBE 100 and 240 $\mu$m data points. The DIRBE calibration gives rise to higher
values of the EBL compared to the FIRAS calibration (see Table 1). Above $\sim$
240 $\mu$m all components were fitted to the FIRAS determination of Fixsen et
al. (1998). 
\end{enumerate}
Finally, we also constructed an "average" EBL spectrum, defined by a polynomial
fit through the nominal 15 $\mu$m lower limit, and the average UV and far--IR
limits. It is approximate in nature, and only used for illustrative purposes.

For sake of brevity the twelve different EBL realizations will be referred to as
XYZ, with X=L, M, or H representing the intensity (low, medium or high) of the
stellar component of the EBL, Y=L or H, representing the low or high intensity
of the 15~$\mu$m EBL flux, and Z=L or H, representing the low or high intensity
of the EBL flux at far--IR wavelengths. So an EBL realization designated as MHL
represents an EBL derived by a polynomial fit through the mid--UV,  the high
mid--IR, and the low far--IR spectral intensities of the EBL.

Figure 2 shows the different realizations of the EBL, and the observational 
data points used in their derivation. The three components are represented in the figure by solid lines (high-UV), lines with connected dots (mid-UV), and dashed lines (low-UV).
Lines going through the high mid-IR are black, whereas those going through the low mid-IR point are grey.
All lines going through the high far--IR data are thick compared to the ones going through the low far--IR data points.
Also shown in the figure is a nominal "average" EBL spectrum which is
represented by a heavy dark line going through the 15 $\mu$m lower limit.
For sake of comparison we also show the range of EBL intensities sampled by the two EBL spectra chosen by de
Jager \& Stecker (2002) and Konopelko et al. (2003) to represent the EBL limits (shaded area in top figure). 
The mid-- and far--IR intensities of these spectra were derived from simple
backward evolution models using highly idealized galaxy spectra (Malkan \&
Stecker 2001). At UV and optical wavelengths, de Jager \& Stecker (2002)
augmented these spectra with the integrated galactic light derived by Madau \&
Pozetti (2000). As is evident from the figure, these two spectra sample a very
small range of viable EBL spectra.  The bottom panel compares our different EBL realizations with the range of EBL spectra derived by Kneiske et al. (2004) for various cosmic star formation histories. Their models do not reproduce the high EBL intensities at UV to mid-IR wavelengths, a common problem in all EBL models. The data points depicted in the figure are
listed in Table~1, and the coefficients of the polynomial approximations are
given in Table 2.

\section{THE $\gamma-$RAY OPACITY OF THE LOCAL UNIVERSE}
The cross section for the $\gamma+\gamma \rightarrow \ e^{+}+e^{-}$ interaction
of a $\gamma$-ray photon of energy $E_{\gamma}$ emitted from a source at
redshift $z$ with a background photon of energy $\epsilon$ is given by (e.g.
Jauch \& Rohrlich 1955)
\begin{eqnarray}
\sigma_{\gamma \gamma}(E_{\gamma},\ \epsilon,\ \mu) & = & {3 \sigma_T\over 16}\
(1-\beta^2)\left[2\beta(\beta^2-2)+(3-\beta^4)\ln\left({1+\beta\over1-\beta}\right)\right]
\\ \nonumber
 \beta & \equiv & \sqrt{1-{\epsilon_{th}\over \epsilon} } \\ \nonumber
 \epsilon_{th}(E_{\gamma},\ \mu) & = & {2(m_ec^2)}^2\over E_{\gamma} (1-\mu) \\
\nonumber
\end{eqnarray}
where $\sigma_T = 6.65\times10^{-25}$ cm$^2$ is the Thompson cross section,
$\epsilon_{th}$ the threshold energy of the interaction, and $\mu \equiv \cos
\theta$, where $\theta$ is the angle between the incident photons. The
$\gamma$-$\gamma$ cross section for the interaction with an isotropic
distribution of background photons has a peak value of 1.70$\times 10^{-25}$
cm$^2$ for $\beta$ = 0.70, which corresponds to energies for which the product
$E_{\gamma}\epsilon \approx 4(m_ec^2)^2 \approx$ 1 MeV$^2$, or
$\lambda_{\epsilon}(\mu{\rm m}) \approx 1.24 E_{\gamma}$(TeV), where
$\lambda_{\epsilon}$ is the wavelength of the background photon.

The optical depth traversed by a photon observed at energy $E_{\gamma}$ that was
emitted by a source at redshift $z$ is given by:
\begin{equation}
\tau_{\gamma}(E_{\gamma},\ z) = \int_0^z \left({{\rm d}\ell \over {\rm
d}z'}\right) {\rm d}z'\int_{-1}^{+1}{\rm d}\mu\ {1-\mu \over
2}\int_{\epsilon'_{th}}^{\infty}{\rm d}\epsilon'\ n_{\epsilon}(\epsilon',\ z')\
\sigma_{\gamma \gamma}(E'_{\gamma},\ \epsilon',\ \mu)
\end{equation}

\noindent
where $n_{\epsilon}(\epsilon',\ z')$d$\epsilon'$ is the comoving number density
of EBL photons with energies between $\epsilon'$ and $\epsilon'+$d$\epsilon'$ at
redshift $z'$,  $\epsilon'_{th}=\epsilon_{th}(E'_{\gamma},\ \mu)$, $E'_{\gamma}
= E_{\gamma}(1+z')$, and where d$\ell/$d$z$, is given by (e.g. Peacock 1999): 
\begin{eqnarray}
\left({{\rm d}\ell \over {\rm d}z}\right) & = & c \left({{\rm d}t \over {\rm
d}z}\right) = {R_H \over (1+z)  E(z)}  \\ \nonumber
E(z) & \equiv & \left\{(1+z)^2 (\Omega_mz+1) + z(2+z)[(1+z)^2 \Omega_r -
\Omega_{\Lambda}]\right\}^{1/2} \qquad ,
\end{eqnarray}
\noindent
where $\Omega_m$ and $\Omega_r$ are, respectively, the matter and radiation
energy density normalized to the critical density,  $\Omega_{\Lambda} =
\Lambda/3H_0^2$ is the dimensionless cosmological constant
($\Omega_m+\Omega_r+\Omega_{\Lambda}$ = 1 in a flat universe), $R_H\equiv c/H_0$
is the Hubble radius, $c$ is the speed of light, and $H_0$ is the Hubble
constant, taken here to be 70~km~s$^{-1}$~Mpc$^{-1}$.
The comoving number density of EBL photons of energy $\epsilon$ at redshift $z$
is given by:
\begin{eqnarray}
\epsilon^2 n_{\epsilon}(\epsilon,\ z) &  = & \left( {4 \pi \over c}\right) \nu
I_{\nu}(\nu,\ z) \\ \nonumber
 &  = & \int_z^{\infty}  \nu' {\cal L}_{\nu}(\nu',\ z') \left| {{\rm d}t \over
{\rm d}z'} \right| { {\rm d}z' \over 1+z'}
\end{eqnarray}
where $\epsilon = h\nu$, $\nu' = \nu (1+z')$, and ${\cal L}_{\nu}(\nu',\ z')$ is
the specific comoving luminosity density at frequency $\nu'$ and redshift $z'$.

Figure 3 (right panel) depicts the TeV opacity of a source located at redshift
$z$ = 0.030 to background photons with an EBL spectrum given by the average
spectrum depicted in the left panel of the figure (see also Figure 2). The
shaded curves in the figure represent the contribution of the different
wavelength regions (depicted in the shaded bar diagram in the left panel) to the
total opacity. The figure illustrates the relation between the rate of increase
in the TeV opacity with $\gamma$-ray energy, and the spectral behavior of the
EBL. Particularly noticable is the decrease in the rise of the opacity between
$\sim$ 1 and 5 TeV, resulting from the dip in the EBL intensity between the
stellar and dust emission peaks.

As an aside, we note that in calculating the $\gamma$-ray opacity we discovered
a numerical error of about 40\% in the polynomial approximations presented by de
Jager \& Stecker (2002) which apparently was caused by the coarse grid used in
the integration of $\tau_{\gamma \gamma}$ over angles (de Jager, private
communication).

\section{NEW GAMMA-RAY DERIVED LIMITS ON THE \\  EXTRAGALACTIC BACKGROUND LIGHT
INTENSITY}

The spectral energy distribution of blazars consists of two spectral components:
 (1) a low-energy X-ray component,
extending up to energies of about 100~keV which is attributed to synchrotron
radiation from energetic electrons, and (2) a high-energy $\gamma$-ray component
with energies extending to
the TeV range, which is usually attributed to inverse Compton (IC) scattering of
the synchrotron emission by energetic electrons (Maraschi, Ghisellini \& Celotti
1992;
Marscher \& Travis 1996).  Competing models exist for the nature of the
particles producing the $\gamma$-ray emission. Leptonic models assume that
energetic electrons are the primary particles producing the $\gamma$-rays by IC
scattering off the synchrotron emission or other ambient soft photons.  Hadronic
models assume that the $\gamma$-ray emission is produced by proton-induced
synchrotron cascades or by decays of secondary particles such as neutral pions
and neutrons (Mannheim \& Biermann 1992; Mannheim 1993, 1998) or, alternatively,
by synchrotron radiating protons (M\"ucke \& Protheroe 2001; Aharonian 2000).

All the above models predict a decline in the blazar luminosity at the highest
$\gamma$-ray energies. Specifically, none of these models predict an exponential
rise of luminosity with energy. 
An exception  is the model of Aharonian, Timokhin, \& Plyasheshnikov (2002b) which was constructed to explain a possible pile-up in the highest energy bins of the intrinsic blazar spectrum. In their model the intrinsic blazar spectrum rises sharply between $\sim$10 and 20 TeV and has an abrupt cutoff at $\sim$20~TeV. 
Coppi (2003) emphasizes that such an upturn in luminosity at TeV energies is not
impossible but unlikely, since it would
require extremely energetic particles in the jet.  Furthermore, the exponential rises produced by some EBL realizations are unbounded and therefore unlikely to be generated by the jet model. Since most blazar models are
moderately successful in explaining the gross features of the low and high
energy emission peaks in the blazar spectra 
over a wide range of energies, we consider any exponential rise in a blazar
spectrum at TeV energies as ``unphysical".

The intrinsic photon spectrum, (d$N$/d$E$)$_i$, of a blazar located at redshift
$z$ is given by:
\begin{equation}
\left({{\rm d}N\over {\rm d}E}\right)_i = \exp[\tau_{\gamma \gamma}(E,\
z)]\times \left({{\rm d}N\over {\rm d}E}\right)_{obs}
\end{equation}
where (d$N$/d$E$)$_{obs}$, the observed spectrum, can be fit with a function
$f_{obs}$ consisting of a power law with an exponential cutoff of the form:
\begin{equation}
f_{obs} = \Phi E^{-\alpha}\times \exp[-E/E_0]
\end{equation}

Using the spectral templates representing the various realizations of the EBL
described in \S3, we calculated the $\gamma$-$\gamma$ opacity towards Mrk~421
and~501, and derived their intrinsic $\gamma$-ray spectra. 
The intrinsic blazar spectrum can be fit by either a parabolic function, $f_p$,
or a parabolic function with an exponential rise, $f_e$, of the form:

\begin{eqnarray}
\left({{\rm d}N \over {\rm d}E}\right)_i &= &  f_p(E,\ \Phi,\ \alpha,\ \beta,) 
\equiv 
                            \Phi\ E^{-\alpha - \beta \log_{10}(E) } \\
                                                           & = &  f_e(E,\ \Phi,\
\alpha,\ \beta,\ E_0)  \equiv  
                            \Phi\ E^{-\alpha - \beta \log_{10}(E) }\times
\exp(E/E_0) 
\end{eqnarray}

\noindent
The function $f_e$ is a generalization of $f_p$, designed to explore intrinsic
spectra that show an ``unphysical" behavior, characterized by a sudden
exponential rise in the function $E^2$(d$N$/d$E$)$_i$ after an initial monotonic
decline or flat behavior with energy. 

We examined the statistical significance of an exponential rise in the intrinsic
blazar spectrum by using the F--test to calculate the probability that the
reduction in the $\chi^2$ of the fit due to the inclusion of the additional
parameter $E_0$ exceeds the value which can be attributed to random fluctuations
in the data. If the reduction in the $\chi^2$ is sufficiently large then the
exponential rise is statistically significant, the intrinsic spectrum is
considered unphysical, and the EBL spectrum causing this behavior is excluded as
a viable realization of the EBL. 

Let $F_{\chi}$ be the ratio $\Delta \chi^2/\chi^2_{\nu_2} \equiv
[\chi^2(\nu_1)-\chi^2(\nu_2)]/\chi^2_{\nu_2}$, where $\chi^2_{\nu_1}$ and
$\chi^2_{\nu_2}$ are the reduced $\chi^2$ for $\nu_1 = N - m$ and $\nu_2 =
N-(m+1)$ degrees of freedom, respectively. The value of $F_{\chi}$ measures the
fractional improvement in $\chi^2$, and its statistics follow that of the
$F(\nu_1,\ \nu_2)~\equiv~(\chi_1^2/\nu_1)/ (\chi_2^2/\nu_2)$ distribution with
$\nu_1=1$, and $\nu_2=N-(m+1)$.  Each $F_{\chi}$ value has therefore an integral
probability distribution $P(F_{\chi},\ \nu_1=1,\ \nu_2)$, which measures the
probability that the improvement in the fit was not a random event. Table 3 list
the resulting probabilities for the F-test on $F_{\chi}$.  A $P$-value of 95\%
or larger is commonly regarded as significant. Consequently, EBL realizations
giving rise to intrinsic spectra for which the exponential rise is significant,
are ruled out as viable spectral representations of the EBL, and are labeled "0"
in the table. Acceptable EBLs are designated with a ``1". The table shows that
Mrk~501 provided more stringent constraints on EBL scenarios. This is not
surprising considering the fact that the $\gamma$-ray observations of this
blazar extend to higher energies and have smaller error bars, compared to the
observations of Mrk~421.

Figures 4 and 5 depict the intrinsic spectra of Mrk~421 and 501, respectively,
for all EBL realizations. The figures clearly show the "unphysical" behavior of
the blazar spectra for the LHH, MHH, MHL, and HHL realizations of the EBL. For
both sources the intrinsic spectra for these realizations initially decrease
with energy, then suddenly increase for at least two data points at energies
above $\sim$ 10 TeV. The rise can be directly attributed to the EBL intensity in
the wavelength regime contributing to the $\gamma$-ray opacity.  All, except
two, of the EBL realization with a high mid-IR intensity give rise  to
unphysical blazar spectra, and are therefore ruled out by the $\gamma$-ray
observation. Of the two viable EBL realizations with high mid-IR intensity, LHL
has a low UV and low far--IR intensity, and HHH has high UV and high far--IR
intensities. So the common characteristic of these two spectra is that they do
not connect between opposite extremes in the UV and far--IR intensities of the
EBL, that is, high UV to low far--IR and vice versa. 

Using only visual inspection of the intrinsic blazar spectra (e.g. Figure 5) ,
one would be tempted to regard additional realizations such as HHH, LLH, MLH,
and LHL unphysical, since they also give rise to an increasing $\gamma$-ray
spectrum. However, the rises occur only at the last datum point, and are
statistically insignificant. This fact is confirmed by the results of the
statistical F-test that show that the confidence level of these rising spectra
is not high enough to ascertain their reality. These examples illustrate the
importance of conducting rigorous statistical tests over casual visual
inspections, for determining the reliability of features in the derived
intrinsic spectrum of blazars.

In addition to the systematic uncertainties of the individual spectral
data points we also considered the possibility that the reported gamma-ray
energies are
higher than their actual values, within the claimed experimental
uncertainties. Use of the published higher energy values would result in an
overestimate of the gamma-ray
opacity and could cause an artificial rise in the absorption corrected spectra.
We have adopted the conservative assumption that the reported energy spectra of
Mrk 501 and Mrk 421 are shifted up in energy by 15\%. We therefore recreated the
observed HEGRA spectra for these sources
 by adopting an energy grid that is 0.85 times the reported one, and derived our
constraints
for the EBL for these spectra using the previously described F-test. The results
show that three previously rejected EBL realizations: LHH, MHL, and HHL are now
viable, and only the MHH one can be firmly rejected.

Figure 6 depicts the effect of the energy shift on the intrinsic spectrum of
Mrk~501. The open squares represent the nominal blazar spectrum, and the solid
ones represent the blazar spectrum when the photon energies are shifted down by
15\%. The heavy dashed and solid lines represent functional fits (a power-law
with an exponential cutoff, eq. 6) through the data. The intrinsic spectra for
both cases are calculated using the LHH realization of the EBL and are, depicted by open 
and solid circles for the nominal and energy-shifted blazar spectrum, respectively. The dashed line is a
functional fit through the intrinsic spectra of the blazar using its nominal
observed spectrum (see also Fig. 5a). As mentioned before, the exponential rise
is statistically significant, and the LHH realization of the EBL was rejected.
The two solid lines represent parabolic and exponentially rising fits to the
intrinsic blazar spectrum, derived from its energy shifted spectrum. The figure
visually shows that the exponentially rising function does not provide a
statistically better fit to the intrinsic blazar spectrum than the parabolic
one. Hence, the LHH realization of the EBL cannot be rejected in this case.   
These results clearly demonstrate that the absolute calibration of atmospheric
Cherenkov telescopes directly impacts
 the ability to  constrain the EBL with $\gamma$-ray astronomy. It is therefore
important that systematic uncertainties be improved in the next generation of
telescopes.

Figure 7 depicts the eight viable EBL spectra (HHH, LLH, MLH, HLH, LHL, LLL,
MLL, and HLL) that give rise to physically ``well behaved" intrinsic
$\gamma$-ray spectra. In addition we show as dashed lines, the three EBL
realizations (LHH, MHL, and HHL) that are viable when the observed $\gamma$-ray
energies are shifted down by 15\%. 
Figure 8 shows the optical depths derived for these EBL realizations. All
optical depths are characterized by a flattening in the 1 to 10 TeV optical
depth caused by the dip in the EBL at mid--IR wavelengths. This is in contrast
to the model of Konopelko et al. (2003) which shows a monotonic increase of
$\gamma$-ray opacity with energy for their adopted EBL spectra. The shaded area
in the figure depicts the optical depth bounded by the two EBL models adopted in
their analysis, and depicted in Figure 2 (upper panel) of this paper. The heavy shaded area represents the range of 
opacities corresponding to the EBL spectra of Kneiske et al. (2004) shown in Figure 2 (lower panel).

From Figure 8 we see that all EBL realizations  imply substantial 
absorption at energies above 400~GeV for blazars with a redshift of more than $
z \approx 0.03$ except for the LLL case (lowest curve in the figure) for which
the spectral shape is only marginally modified by the attenuation. The energy
dependence of the opacity is crucial in determining the shape of the intrinsic
blazar spectrum and the position of its energy peak. This is especially
important for modeling the spectral variability of Mrk~421, for which
observations suggest that its peak energy is shifting to higher TeV energies
with increasing flaring intensity, and for determining the intrinsic spectrum of
H1426+428, which is located at $z$=0.129. 

\section{IMPLICATIONS FOR THE EXTRAGALACTIC \\ BACKGROUND LIGHT INTENSITY}

\subsection{New Limits on the EBL Spectrum}
Using minimal assumptions on the behavior of the intrinsic blazar spectrum we
derived new limits on the spectrum of the EBL. Figure 7 shows the eight template
spectra that did not give rise to an ``unphysical" behavior in the blazar spectrum. 
Table 4 summarizes key characteristics of the eight viable EBL realizations: the
total EBL intensity over the 0.1 to 10$^3$~$\mu$m wavelength range, and the
intensities attributed to starlight (0.1-5~$\mu$m) and reradiated thermal
emission from dust (5-10$^3~\mu$m). The last column gives the fractional
contribution of the dust emission to the total EBL intensity.

Finkbeiner, Davis, \& Schlegel (2000) detected an isotropic emission component
at 60 and 100~$\mu$m after the subtraction of the expected zodiacal and Galactic
emission components from the {\it COBE}/DIRBE maps. The intensity of the
residual emission was found to be 28.1$\pm$7 and
24.6$\pm$8~nW~m$^{-2}$~sr$^{-1}$ at these respective wavelengths. Our analysis
puts a strong upper limit of $\sim$ 15~nW~m$^{-2}$~sr$^{-1}$ on the EBL
intensity at 60~$\mu$m, otherwise the EBL will give rise to an unphysical blazar
spectrum. This upper limit is still within the 2$\sigma$ uncertainty of their
nominally stated detection. However, any isotropic components above this upper
limit must be of local origin. 

The maximum EBL intensity corresponds to the HHH realization is 140~nW m$^{-2}$
sr$^{-1}$, and the minimum EBL intensity, corresponding to LLL, is 48~nW
m$^{-2}$ sr$^{-1}$. The HLL realization gives a fractional dust contribution to
$I_{tot}$ of only 23\%, significantly less than the fractional contribution of
dust to the local luminosity density, which is about 30\% (Soifer \& Neugebauer
1991, Dwek et al. 1998). The fractional contribution of the dust emission to the
EBL intensity should be larger than its contribution to the local luminosity
density, since in the past  galaxies emitted a larger fraction of their total
luminosity at infrared wavelengths, as suggested from the evolution in the
number counts of ultra-luminous IR galaxies with redshift (e.g. Chary \& Elbaz
2001).  So the HLL realization of the EBL may be rejected on spectral grounds.
All other seven EBL realizations give $I_{dust}/I_{tot}$ values larger than
0.34, which is marginally larger than the local value.

\subsection{Local Fluctuations in the Number Density of Background Photons}

The two blazars Mrk~421 and 501 have almost identical redshifts ($z$ = 0.031 and
0.034, respectively), but they may have different opacities if local
fluctuations in the number of infrared galaxies are important.
The optical depth to $\gamma$-ray photons going through an individual disk galaxy
with a mid-IR luminosity $L_{IR}$, mid-IR photon number density $n_{IR}$, and disk radius $R$ is approximately given by:
\begin{eqnarray}
\tau_{gal} & \approx &  \sigma_{\gamma \gamma}\ n_{IR}\ R \\ \nonumber
    & \approx  & \sigma_{\gamma \gamma} \left({1\over c}{L_{IR}/\epsilon_{IR} \over
  R^2} \right)\ R \\ \nonumber
  & \approx  & 0.2\ \sigma_{25} L_{10} R^{-1}_{kpc}
\end{eqnarray}
where $\sigma_{25}$ is the $\gamma$-$\gamma$ cross section in units of
10$^{-25}$ cm$^2$, $L_{10}$ is the mid-IR luminosity of the galaxy in units of
10$^{10}$ L$_{\odot}$, $R_{kpc}$ is a typical galactic radius (in kpc) containing
most of the IR emission from the galaxy, and $\epsilon_{IR}$ is the energy of a mid-IR photon with a wavelength of 10~$\mu$m. Equation (9) shows that for an appropriate combination of $L_{IR}$ and $R$, individual galaxies can be optically thick to TeV $\gamma$-rays. For example, the optical depth will be unity for galaxies with a mid-IR luminosity of $\sim$10$^{12}$~L$_{\odot}$ and a disk radius of $\sim$4~kpc. So a single galaxy along the line of sight to a blazar can have a significant affect on its observed spectrum. However, the probability of encountering such galaxy along  the line of sight to either Mrk~421 or Mrk~501 is very small. The local IR luminosity density is about $2\times10^7$~L$_{\odot}$~Mpc$^{-3}$ (Dwek et al. 1998). Assuming that all this energy is emitted at mid-IR wavelengths gives an upper limit of $n_{gal}\approx10^{-5}$~Mpc$^{-3}$ on the number density of these TeV-thick galaxies. The probability for intersecting such galaxies is given by $P\approx n_{gal}R^2L$, where $L\approx$ 130~Mpc is the distance to $z$ = 0.03. The resulting probability for encountering a galaxy opaque to TeV $\gamma$-rays is about 2$\times 10^{-8}$. It is therefore statistically very unlikely that random fluctuations in the number of galaxies along the line of sight to these two TeV blazars will cause any
significant differences in their TeV opacity.

\section{THE ABSORPTION--CORRECTED BLAZAR SPECTRA }

\subsection{The Intrinsic Average Spectrum of Mrk~421 and Mrk~501}

Figure 9 shows  the absorption--corrected spectra of Mrk~421 applied to the
Whipple and HEGRA $\gamma$-ray data using the eight viable EBL realizations
depicted in Figure 7. The dotted and dashed lines represent the intrinsic blazar
spectra derived by using, respectively, the HLL and LHL realization of the EBL.
The former was rejected on the basis of spectral considerations: the fraction of
the total EBL intensity radiated by dust was too small and inconsistent with the
fact that the infrared luminosity density increases with redshift. The latter,
as can be seen in Figure 5g, gives rise to an unphysical blazar spectrum,
although this unphysical behavior was found to be statistically insignificant. 

We note, that all  absorption--corrected Mrk~421 
spectra exhibit a  curved shape.  Assuming that there is no additional
attenuation by radiation 
fields in the proximity of the blazar jet, these represent the range of possible
intrinsic spectra, from which we can derive their peak luminosity, which
provides important constraints for models for the origin of the $\gamma$-ray
emission. The figures show that  the  
luminosity  peak of Mrk~421 occurs within the energy range between 0.5~TeV 
and 1~TeV. 

Mrk~501 is approximately at the same redshift as Mrk~421, hence EBL absorption
should
affect its spectrum similarly, allowing a direct comparison of the intrinsic
spectra of these two blazars.  
Figure 10 shows absorption--corrected spectra of Mrk~501 for acceptable EBL
realizations. The dotted and dashed line represent the intrinsic spectra for the
same EBL realizations as in Figure 9. The unphysical nature of the LHL-corrected
intrinsic spectrum is clearly shown for the Mrk~501 HEGRA data: the intrinsic
spectrum is essentially flat with energy. 
 The results from HEGRA and Whipple data show that  the intrinsic 
averaged energy spectrum   of Mrk~501 during its flaring activity in 1997 
peaked at energies 
between 0.8 and 2.5~TeV.

A more rigorous determination of the peak energy and its uncertainty can be made
from the parabolic fit to the spectra of these two blazars (equation 7). Since
the uncertainties in the derived peak energies are dominated
by the wide range of EBL realizations, and no truly contemporaneous X-ray data
are available 
for the  complete data set,  this simple function is sufficiently accurate 
to get an estimate of the peak energy of the blazar. 
The use of more elaborate fitting functions based on detailed blazar models may
be justified 
once data obtained with the next generation of atmospheric 
Cerenkov telescopes covering a wider range of $\gamma$-ray energies and
contemporaneous X-ray observations will become available. 

The peak energies, $\rm E_{peak}$, for Mrk~421 and 501 are presented in Table~5.
For Mrk~421 both, the HEGRA 
and the Whipple data sets independently show that the  intrinsic time-averaged
spectrum  of Mrk~421 
during the 2000/2001 season peaked at an energy between 0.5 and 1.2~TeV.
For Mrk~501, the HEGRA data give a consistently higher peak energy than the
Whipple data, ranging from 0.7 to 1.8~TeV for the Whipple, and from 1.1 to
2.5~TeV for the HEGRA data. Furthermore, the table shows that peak energies of
Mrk~501 are systematically higher by about 30 to 60\% compared to that of
Mrk~421 regardless of the EBL realization used to derive their intrinsic
spectra. This point is also illustrated in Figure 11 which depicts the Mrk~501
versus the Mrk~421 peak energies for all acceptable EBL realizations. Table~6
shows the peak energy in the intrinsic spectrum of Mrk~501 obtained by
correcting the energy-shifted spectrum for absorption by the different EBL
realizations. Comparison to the peak energies in Table 5, shows that the general
trend is to shift the peak energies to lower values. Uncertainties in the
determination of the absolute photon energy therefore affects the intrinsic
spectrum and the determination of peak energies. 

The small difference of $\lesssim$ 60\% in the peak energies of these two
blazars while they were in a high flaring state is quite surprising considering
the fact that their historical  synchrotron spectra in X-rays have peaked 
at largely different energies: at $ E >$~100~keV for Mrk~501   (Pian et al.
1998; Catanese et al. 1997), 
and at  $E$ $\sim$ 0.5--8~keV for Mrk~421 (Brinkmann et al. 2000; Krawzcynski et
al. 2001; Tanihata et al. 2004).
The measurements of  the synchrotron peak energy of Mrk~501  were carried out
during the same
months as  the $\gamma$-ray data used in this paper were recorded (see also
Catanese et al. 1997; 
Samuelson et al. 1998). However, the published peak synchrotron energy of
Mrk~421 was not measured contemporaneously with the $\gamma$-ray data.
Spectral measurements at X-ray energies of Mrk~421 during its 2000/2001 flare do
exist, but have not yet been made publicly available.
Preliminary results from this flaring state suggest that the location of the
spectral peak is consistent with those determined in previous years (Krawzcynski
private communication), suggesting that the synchrotron peaks of the two blazars
during their high flaring state are indeed largely different.

\subsection{The Intrinsic Spectral Variability of Mrk~421 }

The measurement of spectral variability of Mrk~421 (Krennrich et al. 2002) allows
further studies of its time-averaged spectra over a 
range of flux levels between 1 and 10~Crab. 
The observed spectra of Mrk~421 show an increased flattening of the spectra with
increasing excitation level, suggesting a systematic shift of the peak energy to
higher energies as a function  of excitation level. To determine the peak energy
of their intrinsic spectrum we corrected the observed spectra for the different
epochs for absorption by all the eight viable EBL realizations derived in \S5. 

Figure 12 (top left panel)  shows the average energy spectra of Mrk~421 of
various flaring states with the 
data binned by flux level based on  measurements with  
the Whipple observatory 10~m $\gamma$-ray telescope.   The remaining panels show the
absorption--corrected flare spectra using the acceptable EBL realizations.
It can be clearly seen that the EBL corrected spectra imply higher peak energies
for the high emission states
than the un-corrected data.  In fact the peak energy for the highest flux level
ranges from
730~GeV (LLL) to approximately 2~TeV (HHL).

Figure 13 depicts the evolution of the peak energy with flaring state. All
energies are normalized to the peak energy of the lowest state, except for the
LHL case for which the peak energies in the first and third flaring state could
not be determined. The figure shows that for all EBL realizations the peak
energy shifts towards higher  energies  with increasing flaring state. Table 7
lists the peak energies for the different flaring states and EBL realizations.
The intrinsic spectrum of Mrk~421 in the lowest 
flaring state has a peak energy that falls significantly below the energy range
covered by the data and therefore
cannot be accurately determined, as evident by the large error bars.  However,
in spite of the uncertainty in the peak energy, the figure does show that for
several EBL realizations, the drop in the peak energy of this flaring state is
significant. Also noticeable in the figure is the relative flattening in the
value of the peak energy between the second and fourth flaring state.
A similar behavior has been observed for Mrk~421 at X-ray energies (George et
al. 1988; Fossati et al. 2000; Brinkmann et  al. 2000) showing low and soft
X-ray states and hard and bright high emission states.

\subsection{The Intrinsic Spectrum of H1426+428 }

A search for other extreme TeV blazars similar to Mrk~501 with synchrotron peak
energies in the hard X-ray band
was carried out by  Costamante et al. (2001) using the BeppoSAX satellite. They
identified
 the BL Lac object  H1426+428 as a strong hard X-ray source.   Spectral
measurements of H1426+428 in 
February 1999  using the LECS, MECS and PDS instruments on BeppoSAX revealed a
peak energy above 
100~keV, similar to Mrk~501 during a long lasting outburst in 1997 (Catanese et
al. 1997; Pian et al. 1998).  
This was followed by the first weak detection of a $\rm \gamma$-ray signal above
300~GeV by Horan 
et al. (2000, 2002).   Consequently, H1426+428 has become an interesting
comparison object to the 
TeV blazars Mrk~421 and Mrk~501.   The third extreme blazar with a high
synchrotron peak energy 
(Giommi, Padovani, \& Perlman 2000)  and shown to emit TeV $\rm \gamma$-ray
radiation (Catanese et al. 1998) 
is  1ES2344+514, but a spectral analysis in $\rm \gamma$-rays  is not available yet.

H1426+428 has a redshift of z=0.129, therefore its $\rm \gamma$-ray spectrum
above 400~GeV is
more strongly attenuated by the EBL than in the case of Mrk~421 and Mrk~501. 
Our results show that for various EBL realizations, the Mrk~421 and Mrk~501
spectra above 400~GeV are already substantially attenuated (see Fig. 8).
  A source  four times more distant should be heavily absorbed above 400~GeV.
As a consequence, a direct comparison of the intrinsic spectrum of H1426+428
with Mrk~421 and
Mrk~501 is more complicated since it depends on the evolution of the EBL in the
$z$=0--0.13 redshift interval.

Having limited the number of EBL realizations using the Mrk~421 and 
Mrk~501 data, we applied the remaining viable EBL realizations to H1426+428
using the spectral 
data from Petry et al. (2002) and Aharonian et al. (2003). In our calculations
we assumed that the EBL is constant in the 0 to 0.13 redshift interval, so that
the $\gamma$-ray opacities given in Figure~8 were simply rescaled to the distance
of H1426+428.
 The absorption-corrected  spectra are shown in Figure 14, and the curves in the
plots represent analytical approximations to the observed and intrinsic source
spectra. Observed as well as intrinsic spectra were fitted by a power law (eq. 6
with $E_0 \rightarrow \infty$), and are shown as solid lines. However, intrinsic
spectra corrected for the HLH, LHL, MLL, and HLL realizations of the EBL were
better fitted by a parabolic function (eq. 7), and are depicted by a dashed line
in the figure.  The figure shows that there is a large spread in peak energies
for the different EBL realizations.  The LLL and LLH realizations yield peak
energies below 400~GeV,  the MLL, HLL, MLH and HLH cases provide  peak energies
of 1-5~TeV whereas the HHH realization suggests a luminosity peak above 10~TeV.   

This range of possible peak energies is significantly different than the one
derived for the $\gamma$-ray spectrum of H1426+428 in previous studies. For
example,  Aharonian et al. (2003) use various theoretical models for the EBL as
well as an extreme phenomenological spectrum to derive an either flat [in
$E^2$(d$N$/d$E$)]
 or a strongly rising intrinsic energy spectrum  of H1426+428. 
Using the Primack et al. (2001) EBL models and an EBL spectrum designed to match
the UV-optical upper limits Costamante et al. (2003) derive an intrinsic
spectrum that peaks at energies above 8-10~TeV.  
Our  work,  using the most current observational limits and detections of the
EBL, suggests alternative possibilities including a strongly rising, an
extremely flat, or a parabolic-shaped intrinsic energy spectrum in the 1-5~TeV
regime. 
Six out of the eight EBL realizations give rise to a soft intrinsic 
spectrum  or a luminosity peak around 1-5~TeV.   Only the extreme HHH and LHL
cases suggests a luminosity peak above 10~TeV.

In fact, three of the EBL scenarios (HLH, MLL, and HLL) yield an intrinsic
energy spectrum for H1426+428 that
exhibits a peak energy of 1-5~TeV similar to  the intrinsic $\rm \gamma$-ray
spectra of 
Mrk~421 and Mrk~501.  
 This concurs well with the fact that Mrk~501 also has a synchrotron peak energy
similar to H1426+428.  

Only  the HHH and LHL realizations lead to  a strongly rising intrinsic spectrum
with energy. These same realizations also yielded an unphysical rise in the
intrinsic spectrum of Mrk~501 (see Fig. 5), but these rises were not considered
to be statistically significant by the F-test.  Consequently,  the question as
to whether or not the TeV $\gamma$-ray spectrum of H1426+428
exhibits a dramatic rise with peak energies above 10~TeV remains unsolved.  
A rising spectrum with a peak energy
above 10~TeV cannot be excluded, but is certainly not favored over the six other
EBL realizations leading 
to peak energies in the few TeV regime or below.

It is clear from the range of absorption-corrected spectra of H1426+428, that
more detailed 
spectral measurements in the $\rm \gamma$-ray regime would be necessary to
resolve the peak energy
of H1426+428 and further narrow the viable range of EBL realizations.
The large systematic error in the  absorption-corrected spectrum results from
the fact that 
H1426+428 is heavily absorbed above 400~GeV and uncertainties in the EBL
realizations enter exponentially
in the derivation of its absorption-corrected spectrum.   The uncertainty in the
EBL spectrum is the
dominant factor in reconstructing the peak energy of the  H1426+428 intrinsic
spectrum.
This is clearly different from the case of Mrk~421 and Mrk~501 for which we were
still able to
provide an estimate of the intrinsic peak energies, despite the large uncertainties
in the EBL.
Blazars like H1426+428 with redshifts of $z >$ 0.1 will be extremely useful for
constraining 
the EBL once measurements by GLAST, HESS, MAGIC  and VERITAS will be available
with sensitivity
starting in the sub-100~GeV 
regime, where the transition from unabsorbed to the absorbed part of the
$\gamma$-ray spectrum 
is expected to occur.  In this region it should again be possible to limit the
EBL cases by 
testing for unphysical rises in the  $\gamma$-ray spectrum.

\section{DISCUSSION AND SUMMARY}

In this paper we presented the observed TeV $\gamma$-ray fluxes from the three
blazars Mrk~421, Mrk~501, and H1426+428, and the evolving spectrum of Mrk~421
during a period of intense variability. The observed blazar fluxes are presented
in Figure 1. The intrinsic spectrum from these sources is attenuated on route to
earth by low energy UV to submillimeter wavelength photons that constitute the
extragalactic background light (EBL). The main objective of this paper was to
explore the range of possible intrinsic source spectra of these nearest blazars,
and derive new limits on the spectrum of the EBL, using the most recent
constraints on the EBL spectrum, and minimal assumptions regarding what
constitutes an unphysical behavior in the blazar spectrum.

Using observed limits and detection of the EBL we constructed a family of twelve
realizations of the EBL spectrum, and derived the intrinsic spectra of the three
blazars. The different spectral templates of the EBL are presented in Figure 2. 
The intrinsic spectra of Mrk~421 and 501 derived for all the EBL realizations
are depicted in Figures 4 and 5, respectively. Some EBL realizations led to an
unphysical behavior in the intrinsic blazar spectrum,  characterized by an
exponential rise after a decline or flat behavior with energy. In \S5 we apply
the F-test to examine the statistical significance of the exponential rise in an
intrinsic blazar spectrum. Table 3 summarizes the results of the F-test,
flagging out EBL realizations that are considered unphysical. 

Discarding EBL realizations that lead to unphysical $\gamma$-ray spectra, leaves
eight viable EBL spectra, shown in Figure~7. Most EBL spectra with high mid--IR
intensity led to unphysical blazar spectra. Table~4 summarizes some
characteristics of these EBL realizations. The range of EBL intensities is
between $\sim$ 50 and 140 nW m$^{-2}$ sr$^{-1}$. The fractional contribution of
dust emission to the total EBL intensity ranges from 0.23 to 0.66. The
contribution of the dust emission to the total luminosity density in the local
universe is about 0.30. Since in the past galaxies reradiated a larger fraction
of their stellar energy output at infrared wavelengths, the spectrum leading to
a fractional dust contribution of 0.23 to the total EBL intensity can be ruled
out as a viable realization of the EBL. 

Figure 8 depicts the $\gamma$-ray opacity of the universe to $z$ = 0.03 for
$H_0$ = 70~km s$^{-1}$ Mpc$^{-1}$. The opacity exhibits a flattening at energies
between 1 and 10~TeV, resulting from the dip in the EBL intensity at mid--IR
wavelengths. The $\gamma$-ray opacities we derive show markedly different
behavior with energy compared to the opacities derived by  Konopelko et al.
(2003), which are monotonically increasing with energy, reflecting  the EBL spectra used in their calculations.

Figures 9 and 10 depict the intrinsic spectra of Mrk~421 and 501 for each of the
eight remaining EBL realizations. Whereas the observed spectra exhibited a
monotonic decrease with $\gamma-$ray energy, all absorption corrected spectra
exhibit a parabolic behavior with energy. This allowed us to identify a peak
energy, $E_{peak}$, in the spectra of these blazars. For all EBL realizations we
found that the peak energy  of Mrk~421 was between 0.5 and 1.2~TeV, whereas that
of Mrk~501 was between 0.8 and 2.5~TeV, depending on the EBL realization. The
value of $E_{peak}$ of Mrk~501 was consistently higher than that of Mrk~421
(Figure 11), but differed by less than 60\%.  X--ray observations of these
blazars showed that their synchrotron peaks are vastly different, typically
$\sim$ 0.4 to 8~keV for Mrk~421, and well above 100~keV for Mrk~501. In light of
their vastly different peak energies in the synchrotron region of their
spectrum, the overall similarity in the value of their peak energies in the
Inverse Compton (TeV) region is extremely interesting. 

The observed spectrum of Mrk~421 during a period of intense spectral variability
is monotonically decreasing with energy at all periods. The absorption corrected
spectrum shows a dramatically different behavior, characterized by the
appearance of a peak in the spectrum that monotonically shifts to higher
energies as the spectrum evolves to higher flux states (Figure 12). The data
suggests a shift in peak energy between the low and high flaring state,
regardless of the EBL spectrum used to correct for the $\gamma$-ray attenuation.
 Figure 13 depict the evolution of $E_{peak}$ with flux level for representative
EBL realizations.

The blazar H1426+428 is the farthest of the blazars detected so far, and its
intrinsic spectrum is therefore most attenuated. The combined effects of the
strong absorption and the uncertainties in its observed spectrum yield an
absorption-corrected spectrum with a wider range of spectral shape than those
suggested in previous investigations. Three of the eight EBL realizations give
rise to an absorption-corrected spectrum that are very similar to the two other
blazars, and are characterized by the appearance of a peak in the $\sim$~1 to
5~TeV energy region (Figure 14). However, two EBL realizations imply a
luminosity peak below 400~GeV, and two cases suggests a peak energy above
10~TeV. More precise knowledge of the spectrum of this blazar, especially
extending the observations into the 100~GeV energy regime will provide important
constraints on the evolution of the EBL with redshift.

We also considered the uncertainties in the absolute calibration of the energy
of $\sim$~TeV~$\gamma$-rays with atmospheric Cherenkov telescopes, and found
that they have an important impact on the ability to constrain the EBL by the
blazar spectrum. The importance of this effect is illustrated in Figure 6, and
described in \S5. Therefore, it will be extremely important for the next
generation of telescopes to improve these systematic uncertainties.

The next generation  imaging telescopes HESS (Hofmann et al. 2003) MAGIC (Lorenz
et al. 2003)  and VERITAS 
(Weekes et al.2002) will cover the sub-100~GeV to 10's TeV energy regime,
covering the transition region between unabsorbed to the strongly  absorbed
region of the spectrum. At energies below $\sim$ 200 GeV EBL absorption is
essentially negligible for a redshift of z=0.03, enabling direct measurements of
the intrinsic spectra of Mrk~421 and Mrk~501 at these energies.  In addition for
the strongest blazars GLAST will help to bridge the 100~GeV regime with
measurements at GeV energies providing an important extension into region where
EBL absorption is negligible.  These measurements will provide additional
constraints on EBL scenarios, since these spectra will have to fit smoothly with
the higher energy intrinsic spectra of these sources that will be heavily
affected by absorption.

Acknowledgement:  We thank Tanja Kneiske for communicating EBL models in digital form, and Martin Schroeder for checking the polynomial coefficients in Table 2. WE also thank the referee, Felix Aharonian, for his useful comments and for pointing out an  error in equation (9) in the original version of the manuscript. ED acknowledges NASA's Long Term Space Astrophysics (LTSA-2003)
Program for support of this work.  FK acknowledges support by the US Department
of Energy.



\begin{deluxetable}{lll}
\tablewidth{0pt}
\tabletypesize {\footnotesize}
\tablecaption{EBL Limits and Detection$^a$}
\tablehead{
\colhead{Wavelength} &
\colhead{ $\nu I_{\nu}$ (nW m$^{-2}$ sr$^{-1}$) } &
\colhead{Reference} 
  }
\startdata 
 0.1595	     &       $>2.9^{+0.6}_{-0.4}$  & Gardner et al. (2000)  \\ 
                       &       $<3.9^{+1.1}_{-0.8}$   & Gardner et al. (2000) \\
   0.2365        &       $<3.6^{+7}_{-0.5}$     & Gardner et al. (2000)  \\
   0.3	 & 12$\pm$7  & Bernstein et al. (2000) \\
    0.36	     &       $2.9^{+0.6}_{-0.4}$  & Madau \& Pozzetti (2000)  \\
      0.45	     &       $4.6^{+0.7}_{-0.5}$  & Madau \& Pozzetti (2000)  \\
   0.555 & 17$\pm$7  & Bernstein et al. (2002) \\
       0.67	     &       $6.7^{+1.3}_{-0.9}$  & Madau \& Pozzetti (2000)  \\
           0.81	     &       $8.0^{+1.6}_{-0.9}$  & Madau \& Pozzetti (2000)  \\
    0.814 & 17$\pm$7  & Bernstein et al. (200*) \\ 
           1.1	     &       $9.7^{+3.0}_{-1.9}$  & Madau \& Pozzetti (2000)  \\
    1.25 & 29$\pm$16$^b$ & Wright (2001)  \\
        & 54$\pm$17$^c$ & Cambr\'esy et al. (2001) \\
               1.6	     &       $9.0^{+2.6}_{-1.7}$  & Madau \& Pozzetti (2000)  \\
                     2.2	     &       $7.9^{+2.0}_{-1.2}$  & Madau \& Pozzetti
(2000)  \\
       & 20$\pm$6$^b$ & Wright (2001)  \\
        & 28$\pm$7$^c$ & Cambr\'esy et al. (2001 \\
     3.5  & 12$\pm$3$^b$ & Wright \& Reese (2000)  \\
        & 16$\pm$4$^c$ & Dwek \& Arendt (1998) \\
      15 & 2.4$\pm$0.5 & Elbaz et al. (2002) \\
       & 2.7$\pm$0.62 & Metcalfe et al. (2003) \\
       24 &$ 2.7^{+1.1}_{-0.7}$ & Papovich et al. (2004) \\
       100  & 23$\pm$6 & Lagache et al. (2000)  \\
              140  & 25$\pm$7$^d$ & Hauser et al. (1998)   \\
                             & 15$\pm$6$^e$ & Hauser et al. (1998)   \\
                 240  & 14$\pm$3$^d$ & Hauser et al. (1998)  \\
                            & 13$\pm$2$^e$ & Hauser et al. (1998)   \\
         240 - 1000 & & Fixsen et al. (1998) \\           
 \enddata
\tablenotetext{ a}{See also Hauser \& Dwek (2001, Tables 1,3, 4) and Arendt \& Dwek (2002, Table 6). Errors represent 1$\sigma$ uncertainties.}
\tablenotetext{b}{Derived using the Wright et al. (2001)
zodiacal light model}
\tablenotetext{c}{Derived using the Kelsall et al.
(1998) zodiacal light model}
\tablenotetext{e}{Based on the DIRBE photometric calibration}
\tablenotetext{f}{Based on the FIRAS photometric calibration}
\end{deluxetable}


 \begin{deluxetable}{lll}
 \tabletypesize{\scriptsize}
 \tablecaption{Polynomial representations for the different EBL spectral
templates$^a$}
 \tablewidth{0pt}
 \tablehead{
 \colhead{Template$^b$} &
 \colhead{polynomial coefficients: $a_j,\  j=0, N$} &
\colhead{$\lambda(\mu$m)-range}
}
 \startdata
 LHH & $\{a\}\equiv$\{0.954270,     0.209657,     -2.03417,      2.72242,     
3.48905,  -9.74560, &  $\lambda > 100 $ \\
   &        3.14729,      6.70213,     -7.21663,      3.01716,    -0.593362,
0.0455855 \}&  \\
 & $\{b\}\equiv$\{0.936188,     0.501445,     -1.43078,    -0.921636,     
1.81965, 0.194080,  & $\lambda < 15$ \\
  &     -0.841639,     0.313789,   -0.0354290\} &  \\
 &
$0.5\times\{\sum_0^{11}a_j[\log_{10}(\lambda)]^j+\sum_0^8b_j[\log_{10}(\lambda)]^j\}
$& 15 $\leq \lambda \leq$ 100  \\
 MHH & \{1.46525,     0.366351,     -2.56385,   -0.0579773,      2.17911,
-0.533745, & all $\lambda$\\
 &     -0.466031,     0.226609,   -0.0276028\} &  \\
 HHH & \{1.78163,    -0.152917,     -4.14983,      4.92401,      3.60609, 
-11.2143, & all $\lambda$\\
  &       4.74911,     5.42408,     -6.51327,      2.78469,    -0.552349,
0.0426156 \}&  \\
 LLH & \{0.969853,     0.603730,     -2.11970,     -2.20808,      2.97575,
1.07215,  &  all $\lambda$\\
   &      -1.92944,     0.670891,   -0.0742040\} &  \\
 MLH &   \{1.47235,     0.175293,     -2.55029,     0.545660,      1.67612,  
-0.954748,  0.139905\} &  $\lambda \leq$ 4 \\
 &   \{1.37639,     0.122876,    -0.136912,    -0.650521,     -5.81963,
5.43507,& $\lambda > 4$  \\
 &           3.92208,     -6.80703,      3.32087,    -0.711689,    0.0578321\} &  \\
HLH  & \{1.58842,      1.31577,     -3.67467,     -3.24647,      5.03638, 1.01891, 
& $\lambda \leq$ 12 \\
  &        -3.03305,      1.34105,    -0.235389,    0.0143138 \}&  \\
&  \{1.89452,     -1.81311,     0.965311,      5.89934,     -15.5846, 5.20984
 & $\lambda > 12$ \\
 &  13.0127,     -14.4141,      6.18153,     -1.23942,    0.0966528\} &  \\
 LHL & \{0.925695,     0.381153,     -1.25192,    -0.160524,      1.11769 
-0.429982, & all $\lambda$\\
   &     0.00277214,    0.0114175 \}&  \\
 MHL & \{1.43906,     0.257308,     -2.22901,     0.432537,      1.41286, 
-0.787225,  0.113213 \}&  all $\lambda$\\
 HHL  & \{1.73013,     0.226485,     -3.33062,      1.07604,      2.24510, -1.71409,
&  $ \lambda \leq $ 2.2 \\
 &         0.425683,   -0.0355017\} &  \\
 & \{1.60828,     0.462253,     -2.66388,     0.294411,      1.77810,
-0.938433,    0.132448\} & $\lambda > 2.2$ \\
LLL & \{0.980669,     0.585033,     -2.41182,     -1.90431,      4.02921, 
-0.175735, & all $\lambda$ \\
   &        -2.44258,      1.93244,    -0.740240,     0.148669,   -0.0123448\} &  \\
MLL & \{1.49096,     0.667023,     -2.85743,     -1.44668,      2.75698,
0.393004, &  all $\lambda$ \\
  &        -1.22980,     0.431022,   -0.0463593\} &  \\
 HLL & \{1.71098,     0.180354,     -2.61936,     0.928249,    -0.298039,
-0.637540, & all  $\lambda$ \\
  &       2.12254     -1.56680     0.450176   -0.0457011\} &  \\
average & \{1.20000,     0.533686,     -1.81116,     -1.28135,      1.94804,
0.655142, &  all $\lambda$ \\
&          -1.19371,     0.405981,   -0.0437552\} &  \\
 \enddata
\tablenotetext{a}{The EBL intensity in nW m$^{-2}$ sr$^{-1}$ is given by: 
$\log_{10}(\nu I_{\nu}(\lambda) = \sum_{j=0}^Na_j[\log_{10}(\lambda)]^j$, \\
where $\lambda$ ranges from 0.1 to 10$^3 \ \mu$m, unless otherwise noted, and
$N$ is the degree of the polynomial.}
\tablenotetext{b}{Templates are abbreviated with L, M, and H. See \S3.2 for a
detailed explanation.}
 \end{deluxetable}

 \begin{deluxetable}{lccccccc}
 \tabletypesize {\footnotesize}
 \tablecaption{Summary of F--Test Results examining the Confidence in the
Exponential Rise in the Intrinsic Blazar Spectra$^a$}
 \tablewidth{0pt}
 \tablehead{
 \colhead{EBL } & \colhead{ } & \colhead{Mrk~421 (HEGRA)} &  \colhead{ } &
\colhead{ } &
                      \colhead{Mrk~501 (HEGRA)}  & \colhead{ } & \colhead{ } \\
 \colhead{Template} &
 \colhead{$\chi_{\nu_1}^2$} &
 \colhead{$\chi_{\nu_2}^2$} &
 \colhead{$P(F_{\chi},\ 1,\ \nu_2)$} &
  \colhead{$\chi_{\nu_1}^2$} &
 \colhead{$\chi_{\nu_2}^2$} &
 \colhead{$P(F_{\chi},\ 1,\ \nu_2)$} &
\colhead{EBL } 
}
 \startdata
 LHH   &  1.877 & 1.366 & 0.88 & 3.18 & 1.84 & 0.97 & 0$^b$ \\
 MHH &2.794 & 1.528 & 0.97 & 5.377 & 1.889 & 0.99 & 0 \\
 HHH & 1.425 & 1.249 & 0.68 & 2.136 & 1.755 & 0.79 & 1 \\
 LLH  & 1.249 & 1.244 & 0.14 & 1.847 & 1.797 & 0.35 & 1 \\
 MLH & 1.333 & 1.339 & 0.0 & 1.895 & 1.819 & 0.43 & 1 \\ 
HLH & 1.616 & 1.096 & 0.91 & 2.240 & 1.726 & 0.85 & 1 \\
 LHL & 1.455 & 1.216 & 0.75 & 1.912 & 1.404 & 0.88 & 1 \\
 MHL &1.616 & 1.236 & 0.84 & 2.194 & 1.389 & 0.95 & 0$^b$ \\
 HHL & 1.760 & 1.196 & 0.91 & 2.34 & 1.34 & 0.97 & 0$^b$ \\
 LLL & 1.036 & 1.111& 0.0 & 1.326 & 1.446 & 0.0 & 1 \\
MLL &1.226 & 1.106 & 0.62 & 1.450 & 1.418 & 0.32 & 1 \\
 HLL & 1.118 & 1.101 & 0.27 & 1.301 & 1.383 & 0.0 & 1$^c$ \\ 
\enddata
\tablenotetext{a}{$F_{\chi}(1,\ \nu_2)$ probabilities were calculated to test
the confidence that the inclusion of an exponential rise in the intrinsic blazar
spectrum improves the $\chi^2$ of the fit beyond that expected from random
fluctuations in the observational data. The confidence limit was set  at a
probability of 95\%. Larger probabilities suggest a high degree of confidence 
that the exponential rise in the intrinsic blazar spectrum is real. EBL
realizations responsible for such rise in the spectrum of either one of the
blazars were considered invalid and labeled "0". Acceptable EBL are labeled as
"1". }
\tablenotetext{b}{These EBL realizations are still viable when the $\gamma$-ray
energy scale is shifted down by 15\%, reflecting the uncertainty in the absolute
photon energy calibration with atmospheric Cerenkov techniques.}
\tablenotetext{c}{The HLL realization can be further excluded from EBL
considerations (see table 6)} 
 \end{deluxetable}

 \begin{deluxetable}{ccccc}
 \tablecaption{Characteristic of the Viable EBL Realizations$^a$}
 \tablewidth{0pt}
 \tablehead{
 \colhead{EBL} & 
 \colhead{$I_{tot}$} & 
 \colhead{$I_{stars}$} & 
 \colhead{$I_{dust}$} & 
 \colhead{$I_{dust}$/$I_{tot}$} \\
 \colhead{realization} & 
 \colhead{(0.1 -- 10$^3 \ \mu$m)} & 
 \colhead{(0.1 -- 5 $\ \mu$m)} & 
 \colhead{(5 -- 10$^3 \ \mu$m)} &
 \colhead{   }  
 }
 \startdata
   HHH & 139.8 & 91.5 & 48.2 & 0.35 \\
   LLH & 58.9 & 20.0 & 38.9 & 0.66 \\
   MLH & 93.6 & 54.0 & 39.6 & 0.42 \\
   HLH & 108.5 & 70.1 & 38.4 & 0.35 \\
   LHL & 58.4 & 21.1 & 37.3 & 0.64 \\
  LLL & 48.1 & 20.1 & 28.0 & 0.58 \\ 
  MLL & 84.5 & 56.0 & 28.5 & 0.34 \\
  HLL & 112.8 & 87.3 & 25.5 & 0.23 \\
 \enddata
 \tablenotetext{a}{All EBL intensities are given in units of nW m$^{-2}$
sr$^{-1}$. All EBL realizations listed above gave rise to physically
"well-behaved" intrinsic blazar spectra. The EBL realization HLL can be further
excluded because its $I_{dust}$/$I_{tot}$ ratio is lower than that in the local
universe. See text for further details.}
 \end{deluxetable}

%

 \begin{deluxetable}{lccccccc}
 \tablecaption{Peak energies $\rm E_{peak}(GeV)$ of absorption--corrected 
energy spectra of Mrk~421 and Mrk~501.}
 \tablehead{
 \colhead{EBL } & \colhead{ } &  \multicolumn{2}{c}{Mrk~421} & \colhead{ } &
\colhead{ }  & \multicolumn{2}{c}{Mrk~501} \\
\colhead{Mrk 501} \\
 \colhead{template} & \colhead{ } &
 \colhead{Whipple} & \colhead{HEGRA} & \colhead{ } &  \colhead{ } &
\colhead{Whipple} & \colhead{HEGRA}  
 }
 \startdata
  HHH  &  & $\rm 1200 \: \pm \: 60 $  &  $\rm 1241 \: \pm \: 84 $ &  &  &$\rm
1778 \: \pm \: 185 $ & $\rm 2460 \:  \pm \: 142 $   \\
 LLH  & & $\rm 450 \:  \pm \: 51 $  &  $\rm 800 \:  \pm \: 79 $  &  &  & $\rm
773 \: \pm \: 159 $ & $\rm 1100 \:  \pm \: 133 $   \\
  MLH  &  & $\rm 777 \:  \pm \: 51 $  &  $\rm 1048 \:  \pm \: 72 $  & & & $\rm
1136 \: \pm \: 142 $ & $\rm 1543 \:  \pm \: 94 $   \\
 HLH  &  & $\rm 1024 \:  \pm \: 50 $  &   $\rm 1144 \:  \pm \: 65 $ & & & $\rm
1348 \: \pm \: 107 $ & $\rm 1640 \:  \pm \: 75 $ \\
 LHL   & & 331$\pm$87 & 790$\pm$107 & & & 1030 $\pm$249 & \nodata \\
 LLL  & & $\rm 468 \:  \pm \: 52 $  &  $\rm 818 \: \pm \: 75 $ & & & $\rm 778 \:
\pm \: 155 $ & $\rm 1172 \:  \pm \: 118 $  \\ 
 MLL  & & $\rm 837 \:  \pm \: 52 $  &  $\rm 1043 \: \pm \: 70 $  & & & $\rm 1190
\: \pm \: 138 $ & $\rm 1546 \:  \pm \: 89 $   \\ 
 HLL  & & $\rm 999 \:  \pm \: 47 $  &  $\rm 1151 \: \pm \: 62 $  & & & $\rm 1321
\: \pm \: 123 $ & $\rm 1638 \:  \pm \: 70 $  \\ 
 \enddata
 \end{deluxetable}


 \begin{deluxetable}{lcccc}
 \tablecaption{Peak energies $\rm E_{peak}(GeV)$ of absorption--corrected HEGRA
 energy spectra of Mrk~421 and Mrk~501. The observed photon energies were
shifted down by 15\%.}
 \tablehead{
 \colhead{EBL Template} & \colhead{ } & \colhead{Mrk~421} & \colhead{ } &
\colhead{Mrk~501} 
}
 \startdata
 LHH & & 697 $\pm$ 81 & & 1615 $\pm$340 \\
  HHH  & &1130$\pm$70 & &  2109$\pm$ 103   \\
 LLH  & &729 $\pm$65 & & 1055$\pm$98   \\
  MLH  & & 960$\pm$60 & & 1431 $\pm$74   \\
 HLH  & & 1081$\pm$52 & & 1559$\pm$60 \\
 LHL   & & 712$\pm$85 & & 1969$\pm$360 \\
 MHL & & 981$\pm$77 & & 2131$\pm$152 \\
 HHL & & 1156$\pm$68 & & 2082$\pm$96 \\
 LLL  & & 736$\pm$63 & & 1088$\pm$88  \\ 
 MLL  & & 969$\pm$59 &  & 1448$\pm$70 \\ 
 HLL  & & 1075$\pm$55 & & 1544$\pm$57  \\ 
 \enddata
 \end{deluxetable}

 \begin{deluxetable}{lcccccc}
 \tablecaption{The Evolution of Peak Energies (GeV) in the Mrk~421 Spectrum 
with Flaring State for Different EBL Realizations}
 \tablewidth{0pt}
 \tablehead{
 \colhead{EBL } & \colhead{ } & \colhead{ } &  \colhead{ Flaring} &
\colhead{State } &
                      \colhead{ }   & \colhead{ }  \\
 \colhead{Realization} &
 \colhead{1} &
 \colhead{2} &
 \colhead{3} &
  \colhead{4} &
 \colhead{5} &
 \colhead{6} 
}
 \startdata
 HHH & 85$\pm$318 & 1055$\pm$303 & 1105$\pm$205 & 1340$\pm$197 & 1490$\pm$131 &
2078$\pm$290  \\
 LLH  &  31$\pm$90 & 448$\pm$245 & 139$\pm$123 & 581$\pm$165 & 613$\pm$111 &
725$\pm$258  \\
 MLH & 4$\pm$88 & 744$\pm$250 & 608$\pm$148 & 925$\pm$159 & 995$\pm$104 &
1230$\pm$213  \\ 
 HLH  & 332$\pm$312 & 986$\pm$238 & 978$\pm$133 & 1158$\pm$145 & 1231$\pm$92 &
1457$\pm$173  \\
 LHL  &     \nodata           &  362$\pm$321  &         \nodata       & 
539$\pm$263  & 583$\pm$208 & 1345$\pm$1952  \\
 LLL  & 175 $\pm$172  &  446 $\pm$244   &  143 $\pm$123  &  582 $\pm$164  &  615
$\pm$110  &  730 $\pm$254  \\
MLL  &  47 $\pm$197   &  806 $\pm$248   &  706 $\pm$145  &  982 $\pm$156  & 
1054 $\pm$101  &  1288 $\pm$203   \\
 HLL & 284 $\pm$320 & 960 $\pm$237 & 942 $\pm$133 & 1133 $\pm$145 & 1205 $\pm$92
& 1430 $\pm$174   \\ 
\enddata
\tablenotetext{a}{ Energies in units of TeV }
 \end{deluxetable}

\clearpage

\begin{figure}[t]
\plotone{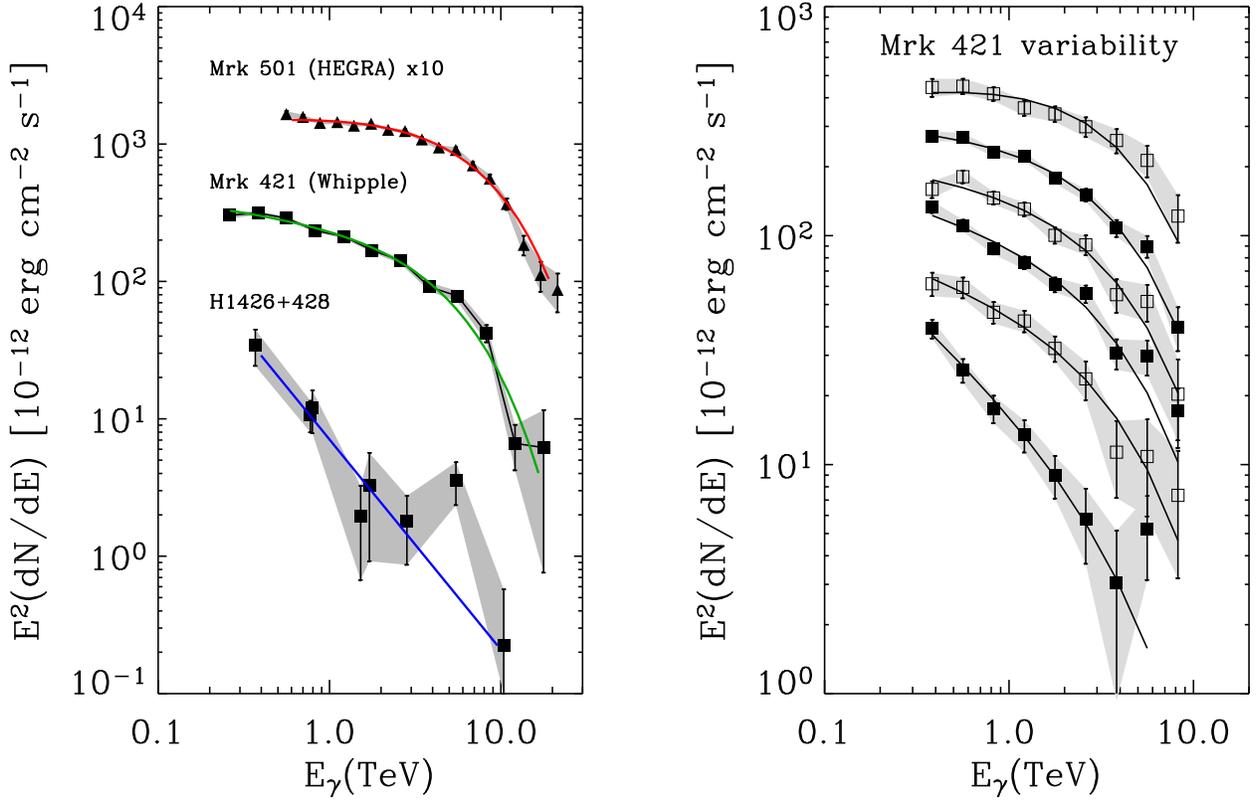}
 \caption{The observed $\gamma-$ray spectra of Mrk~421, Mrk~501, and H1426+428
(left panel). The right panel depicts the spectral variability of Mrk~421. The
lines are analytical fits represented by a power law with an exponential cutoff
(see eq. 6) to the data. See \S2 for more details.}
\end{figure}

\begin{figure}[t]
\epsscale{0.6}
\plotone{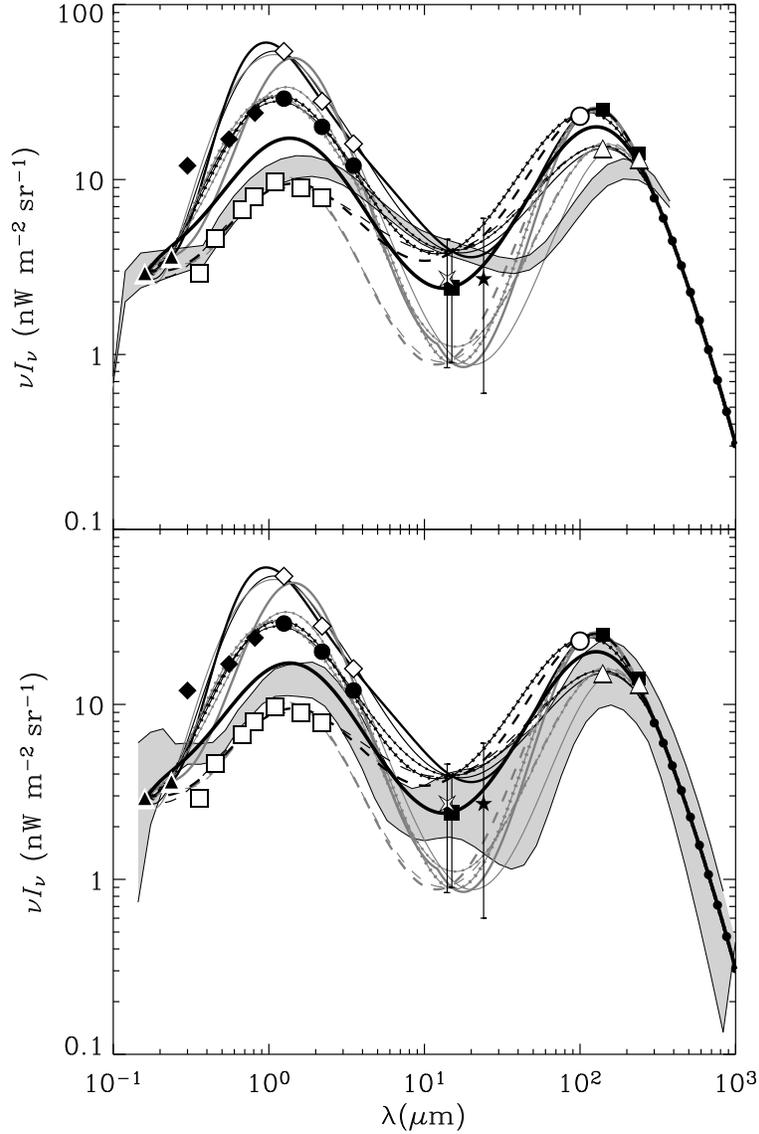}
\epsscale{1.0}
  \caption{Template spectra representing different realizations of the EBL with
select limits and detections of the EBL: (1) black triangles--Gardner et al.
(2000); (2) white squares--Madau \& Pozzetti (2000); (3) black
diamonds--Bernstein et al. (2002); (4) black circles--Wright (2001), Wright \&
Reese (2000); (5) white diamonds--Cambr\'esy et al. (2001); (6) black
square--Elbaz et al. (2002); (7) white star--Metcalfe et al. (2003); (8) black star--Papovich et al. (2004); (9)  white circle--Lagache et al. (2000); (10) black
squares--Hauser et al. (1998, DIRBE calibration); (11) white triangles--Hauser et
al. (1998, FIRAS calibration); (12) small black circles: Fixsen et al. (1998).
The data and the uncertainties are summarized in Table 1. The heavy dashed line
going through the 15~$\mu$m point represents an ``average" EBL spectrum, used
here only for illustrative purposes. The shaded area in the top figure is bounded by the two EBL
model spectra  presented in de~Jager \& Stecker (2002), and the one in the bottom figure represents the range of EBL intensities presented by Kneiske et al. (2004).}
  \end{figure}

\begin{figure}[t]
\plotone{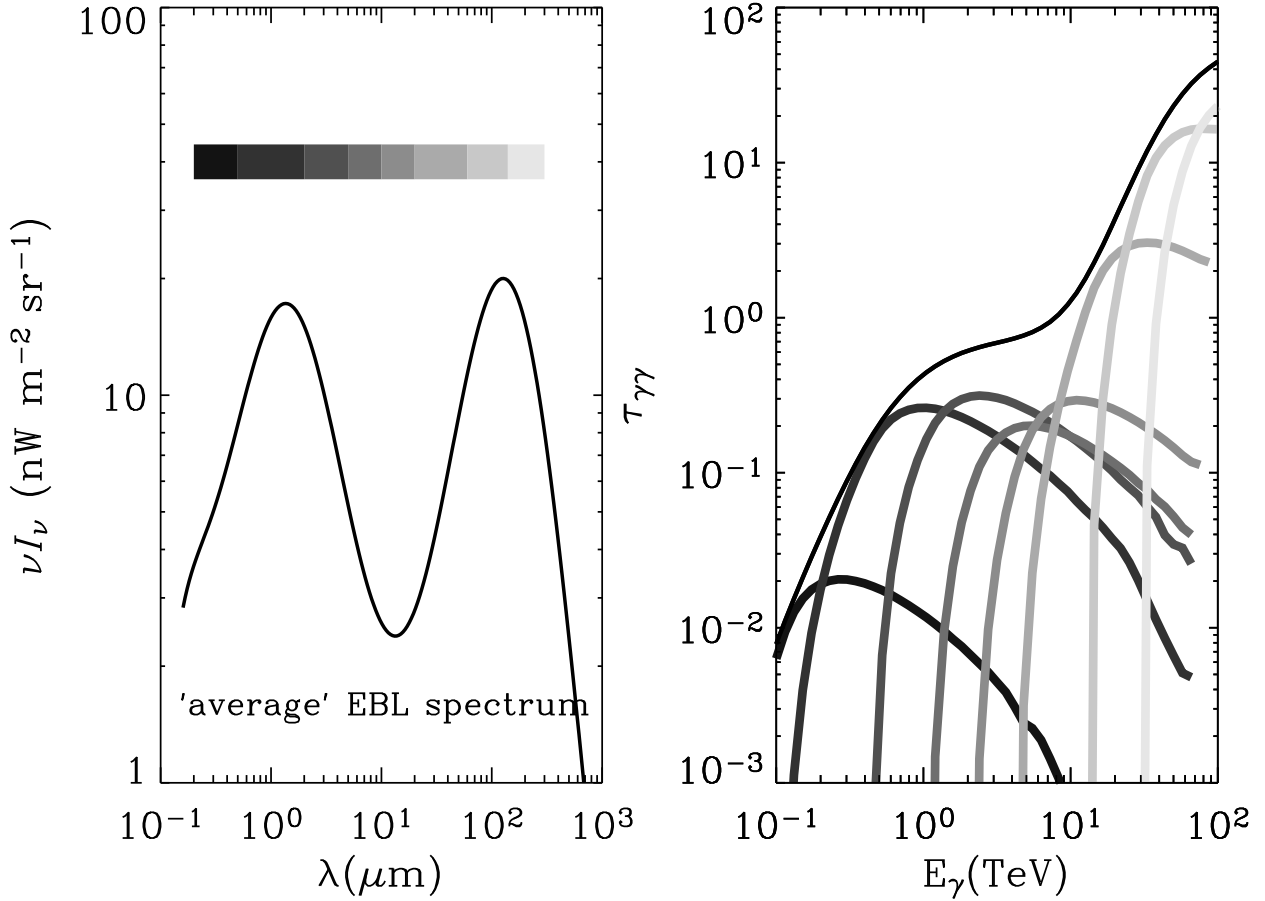}
  \caption{ {\it Left panel}: The 'average' EBL spectrum as defined in \S3 in
the text. The shaded bar diagram indicates the different wavelength regions
depicted in the right panel of the figure. {\it Right panel}: The $\gamma$-ray
opacity of a source located at redshift $z$=0.03 is shown as a thin black line.
The shaded curves in the figure represent the contributions of the different
wavelength regions (depicted in the left panel) to the total opacity. A value of
$H_0$ = 70~km~s$^{-1}$~Mpc$^{-1}$ was used in the calculations.}
  \end{figure}

\begin{figure}[t]
\begin{center}
\plotone{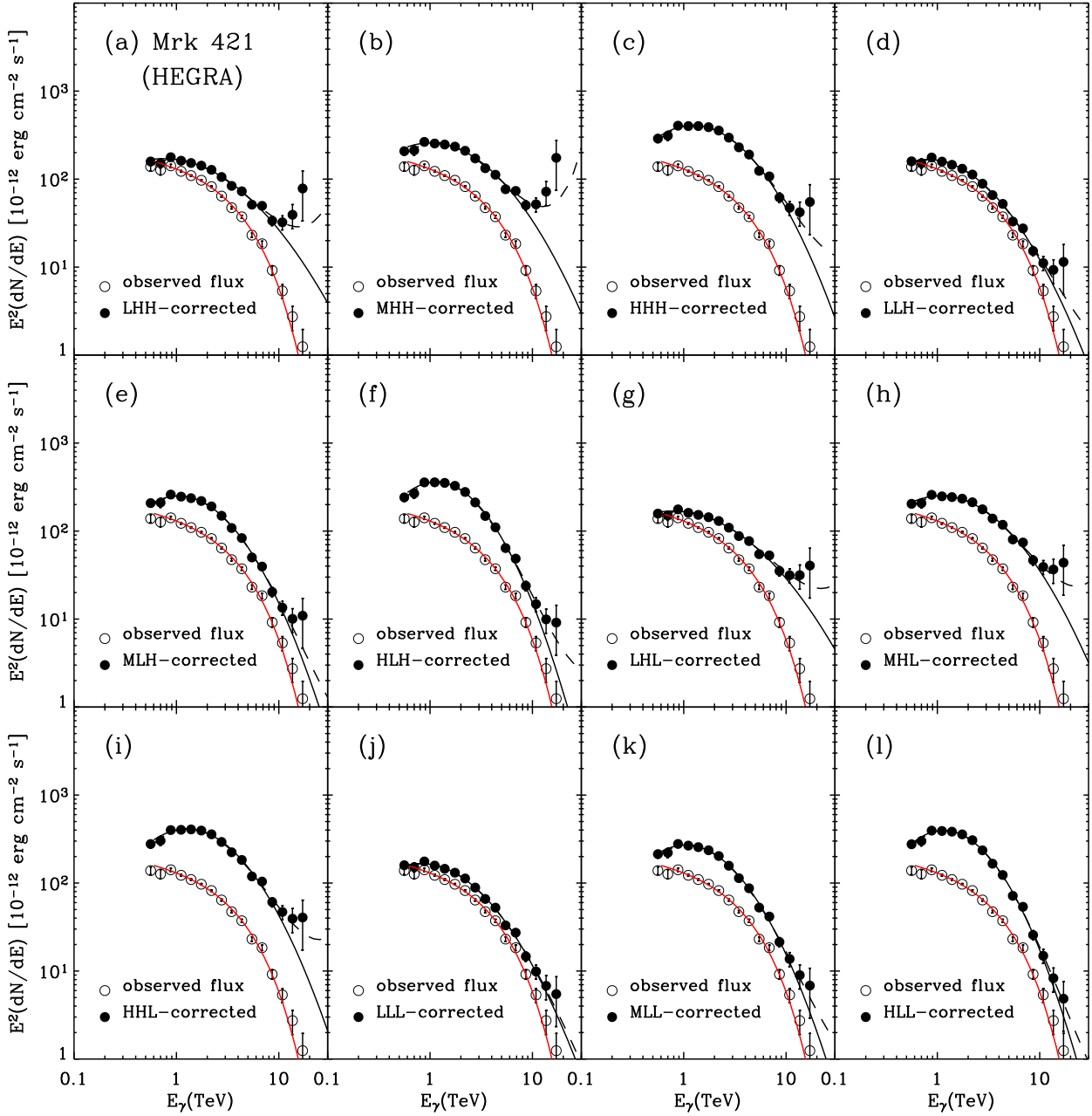}
  \end{center}
  \caption{The measured energy spectra of Mrk~421 (Aharonian et al. 2002a)  and
the absorption--corrected spectra  using all the twelve EBL realization depicted
in Figure 2 are plotted versus energy. Also shown are fits to the observations
(open circles) given by a power law with an exponential cutoff (eq. 6); a
parabolic fit (solid curve, eq. 7) to the intrinsic spectrum (filled circles);
and a parabolic fit with an exponential rise (dashed curve, eq. 8) to the
intrinsic spectrum.
 }
  \end{figure}

\begin{figure}[t]
\begin{center}
\plotone{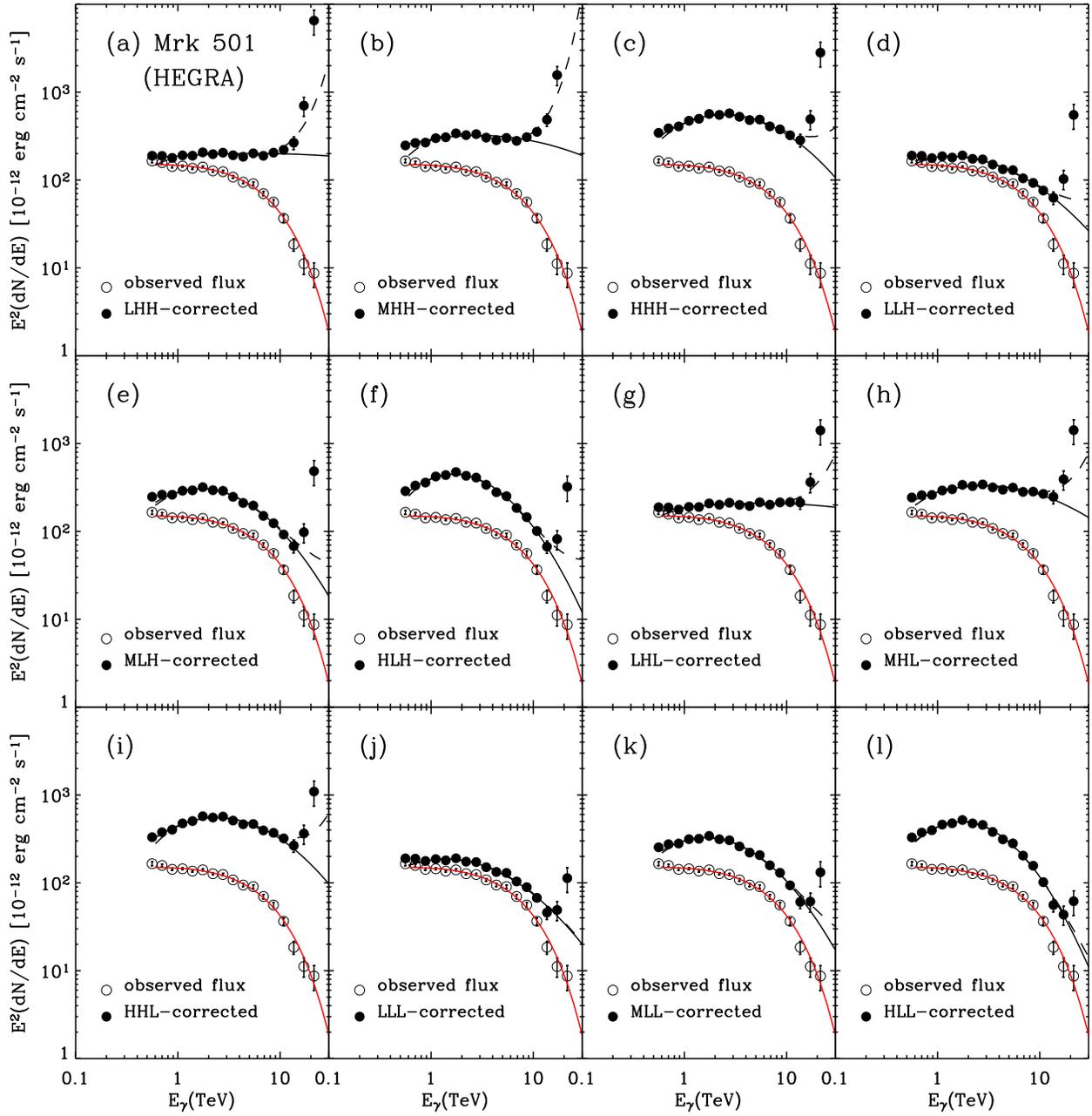}
  \end{center}
  \caption{Same as Figure 4 for Mrk~501 data (Aharonian et al. 1999)}
  \end{figure}

\begin{figure}[t]
\epsscale{0.7}
\plotone{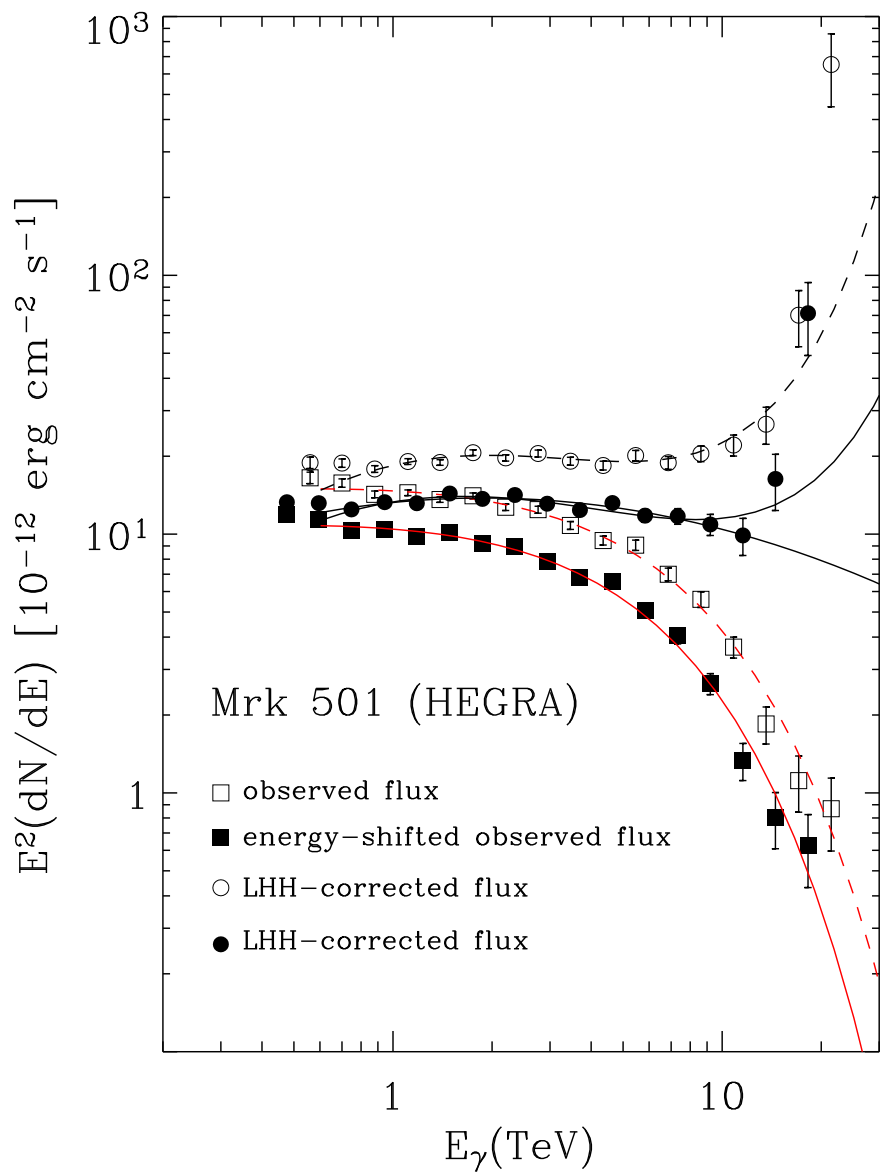}
\epsscale{1.0}
  \caption{The effect of the 15\% shift in the energy of the photons on the
intrinsic blazar spectrum, and on the ability to reject specific EBL
realizations. The figure shows that the LHH realization that was previously
rejected is now viable when the blazar energy is shifted down by 15\%. More
details in \S5 of the text.}
  \end{figure}

\begin{figure}[t]
\plotone{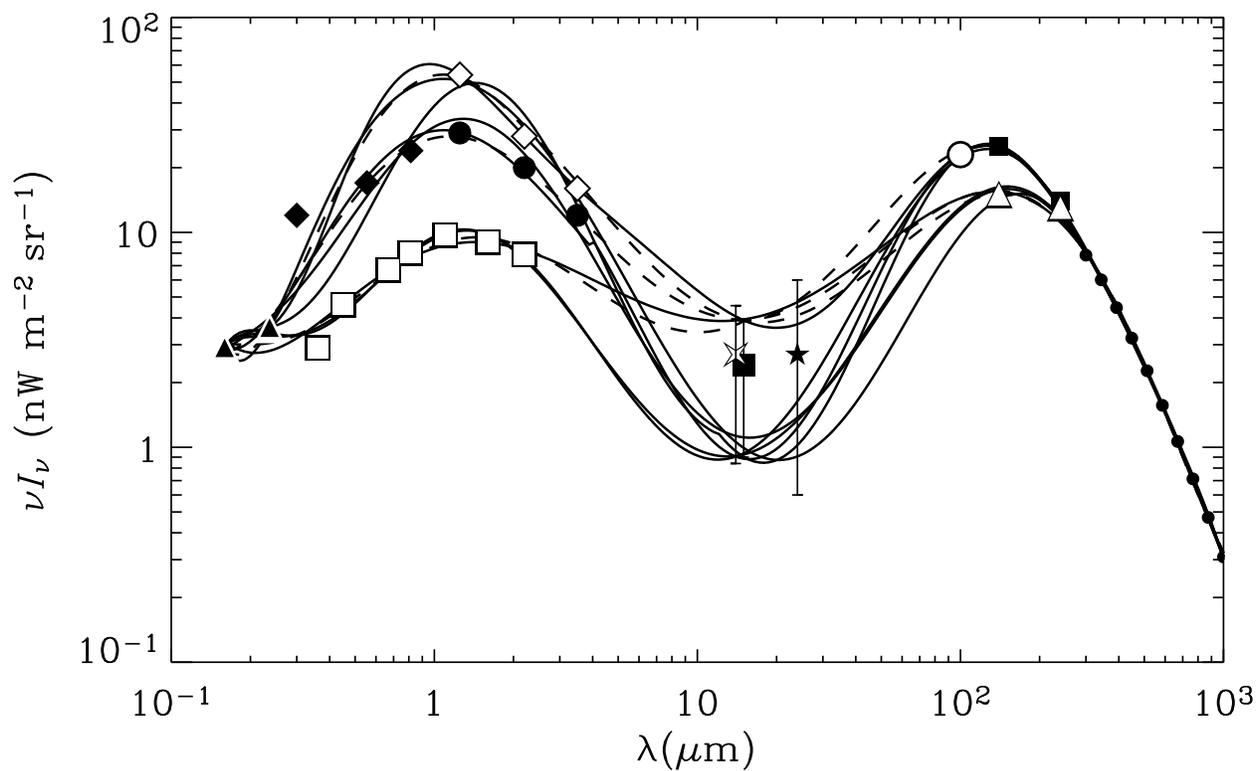}
  \caption{The eight EBL realizations  that do not give rise to an unphysical
intrinsic blazar spectrum (see Table 3) are shown in solid lines. Dashed lines
represent the three previously rejected EBL realization (LHH, MHL, and HHL) that
become viable when the $\gamma$-ray energies are shifted down by 15\%. Symbols
for the observational limits and detections are the same as in Figure 2.}
  \end{figure}

\begin{figure}[t]
\plotone{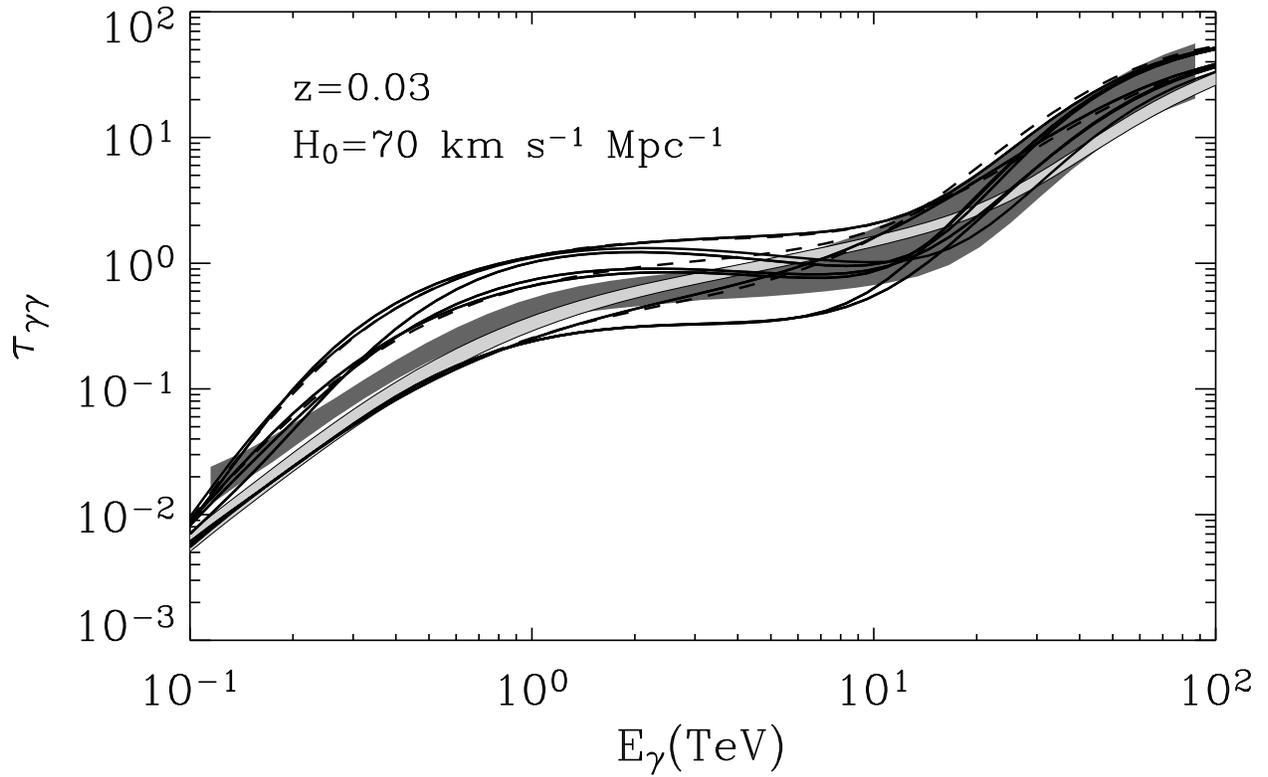}
  \caption{The $\gamma$-ray opacity for the eight EBL spectra depicted in Figure
6. The lightly shaded area in the figure is bounded by the opacities corresponding to
the two EBL spectra adopted by Konopelko et al. (2003), and the darkly shaded one by those adopted by Kneiske et al. (2004).}
  \end{figure}

\begin{figure}[t]
\epsscale{0.6}
\begin{center}
\plotone{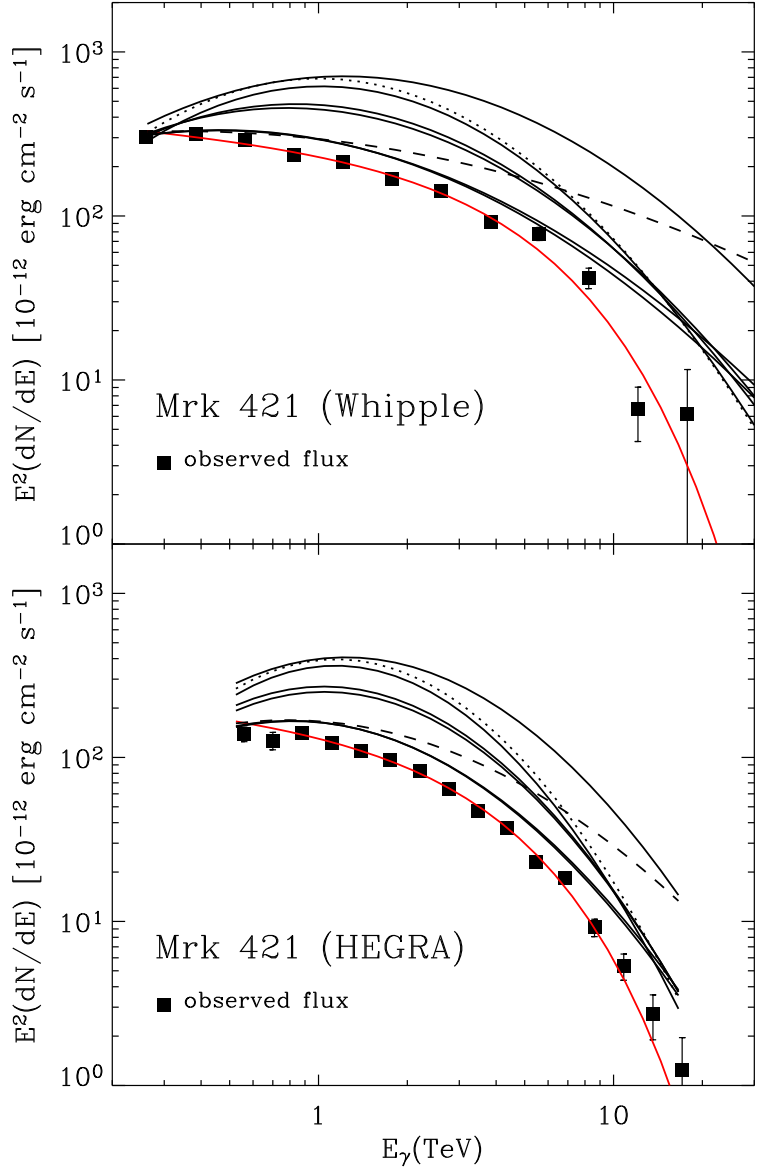}
\epsscale{1.0}
  \end{center}
  \caption{Top panel: The measured energy spectrum of Mrk~421  in 2000/2001,
based on the Whipple data (Krennrich et al. 2001) is plotted versus energy. The
solid line through the observed spectrum (filled squares) is an analytical fit
characterized by a power law with an exponential cutoff. The other curves depict
the absorption corrected spectra for the eight viable EBL realizations. Of these
the HLL realization (dotted line) can be rejected on spectral grounds (see
discussion in \S6.1). The LHL realization (dashed line) gives rise to an
unphysical intrinsic spectrum, which is more evident in Figure 8, but cannot be
rigorously ruled out by the F-test. Bottom panel: the same as the top panel for
the HEGRA data (Aharonian et al. 2002a).}
  \end{figure}

\begin{figure}[t]
\epsscale{0.7}
\begin{center}
\plotone{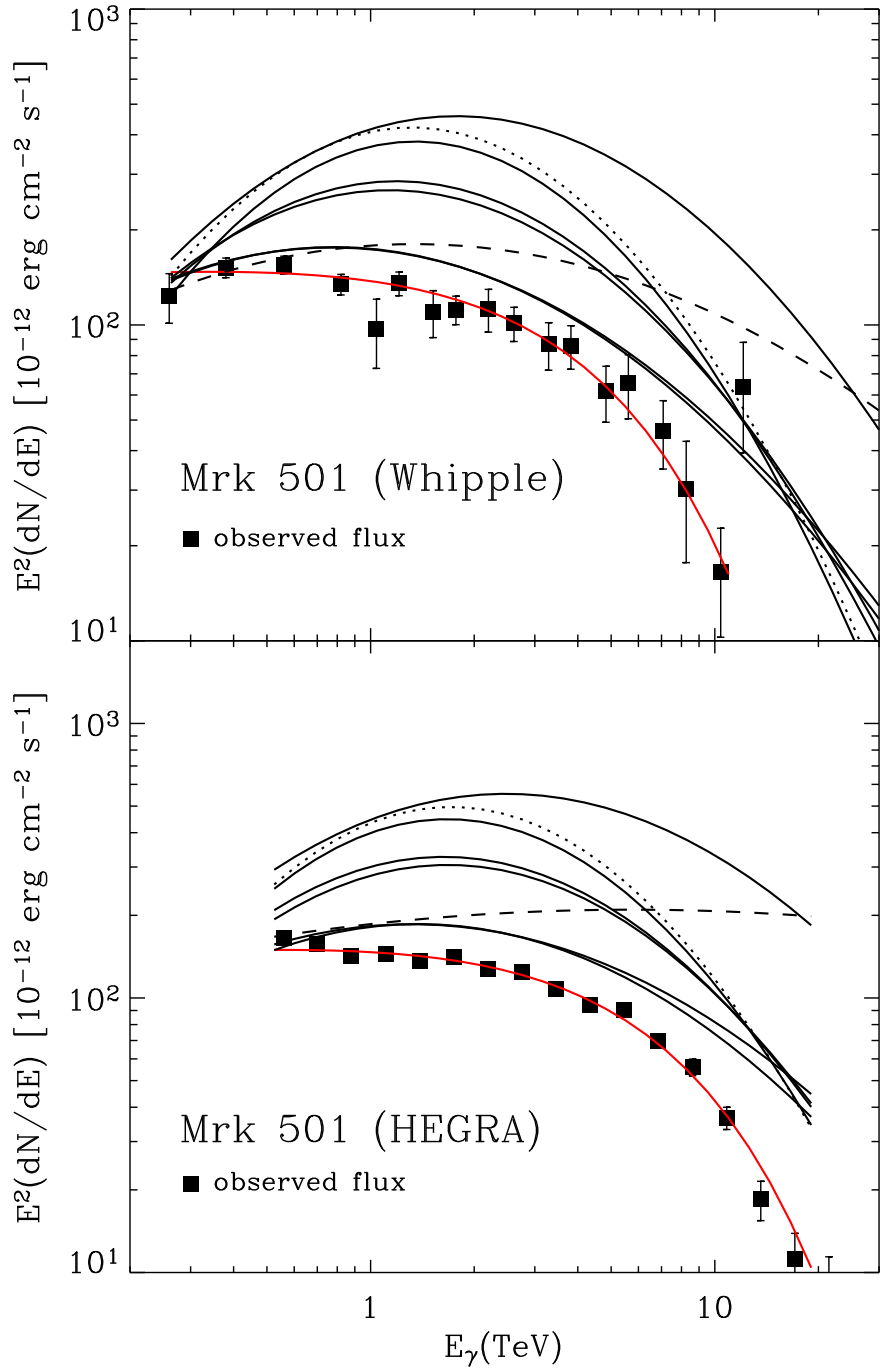}
\epsscale{1.0}
  \end{center}
  \caption{same as Figure 8 for the Mrk~501 data The Whipple data are from
(Samuelson et al. 1998) and the HEGRA from (Aharonian et al. 1999).}
  \end{figure}

\begin{figure}[t]
\plotone{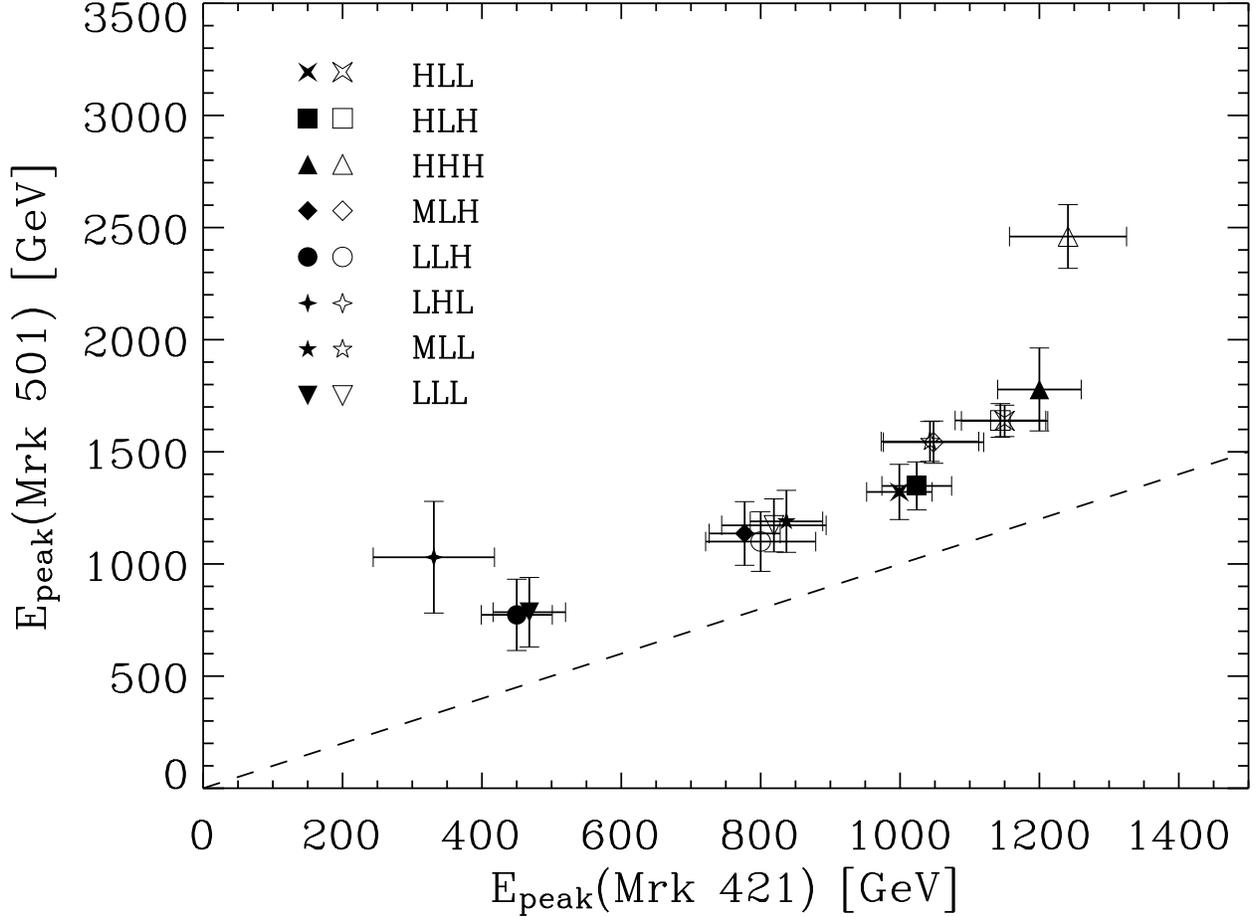}
  \caption{The peaks of the intrinsic $\gamma$-ray spectrum of Mrk~421 derived
for the eight viable EBL realizations are plotted against the same quantities
for Mrk~501. The different symbols represent the EBL templates used in deriving
the intrinsic blazar spectra from the Whipple observations (filled symbols) or
the HEGRA observations (open symbols). The dashed line represents the points
where $E_{peak}$(Mrk~421) = $E_{peak}$(Mrk~501). The figure shows that peak
energies for Mrk~501 are systematically higher than those for Mrk~421.}
  \end{figure}

\begin{figure}[t]
\begin{center}
\epsscale{0.8}
\plotone{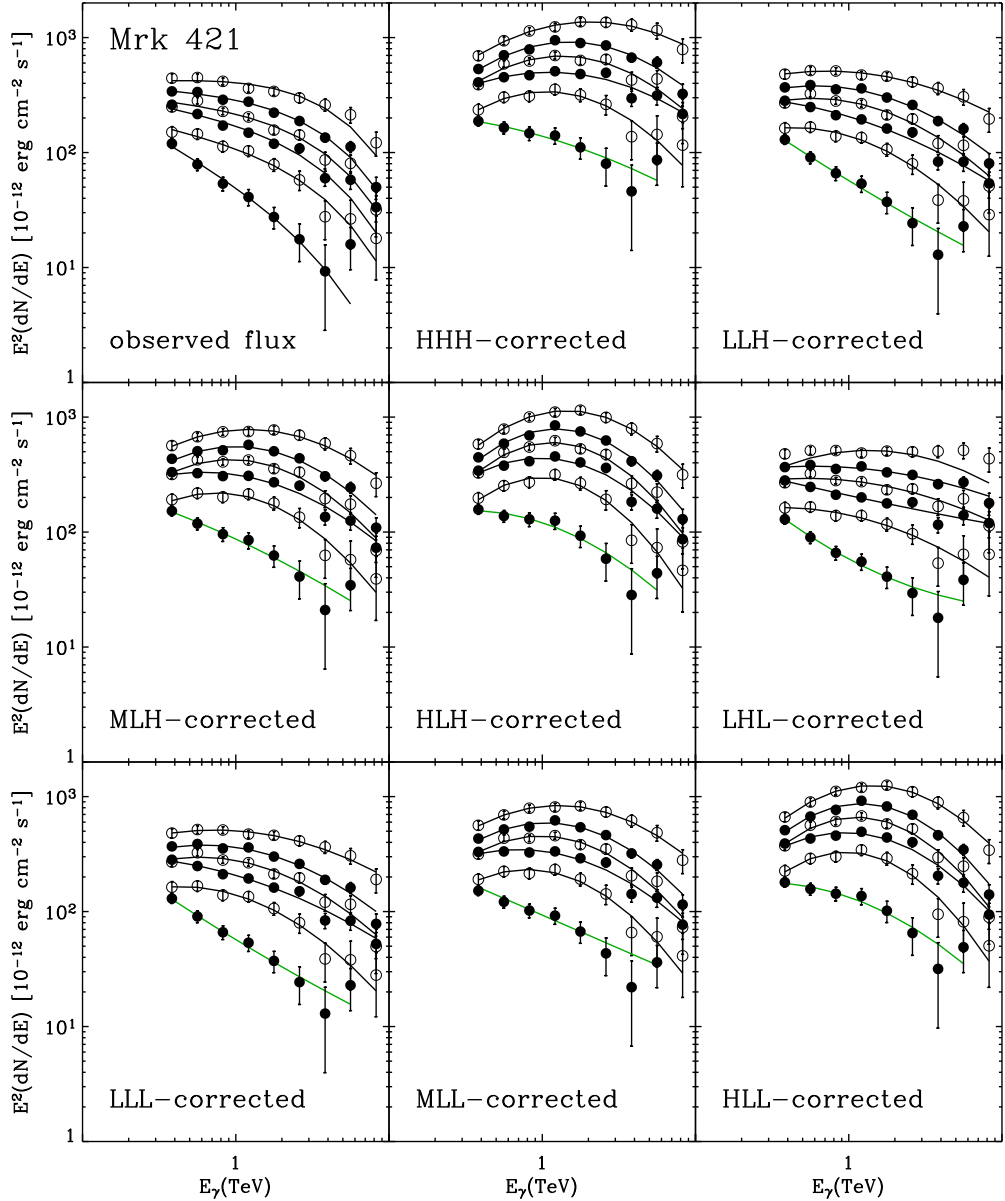}
\epsscale{1.0}
  \end{center}
  \caption{The observed spectrum of Mrk~421 during various stages of flaring
activity is plotted versus energy in the top left panel of the figure. The other
panels show the absorption-corrected spectra for the eight viable EBL scenarios
listed in Table 3. The HLL realization of the EBL can be rejected because of the
low fractional contribution of the dust emission to the total EBL intensity
(\S6.1), and the LHL realization yields an unphysical intrinsic blazar spectrum
which, however, cannot be rejected by the F-test. }
  \end{figure}

\begin{figure}[t]
\begin{center}
\plotone{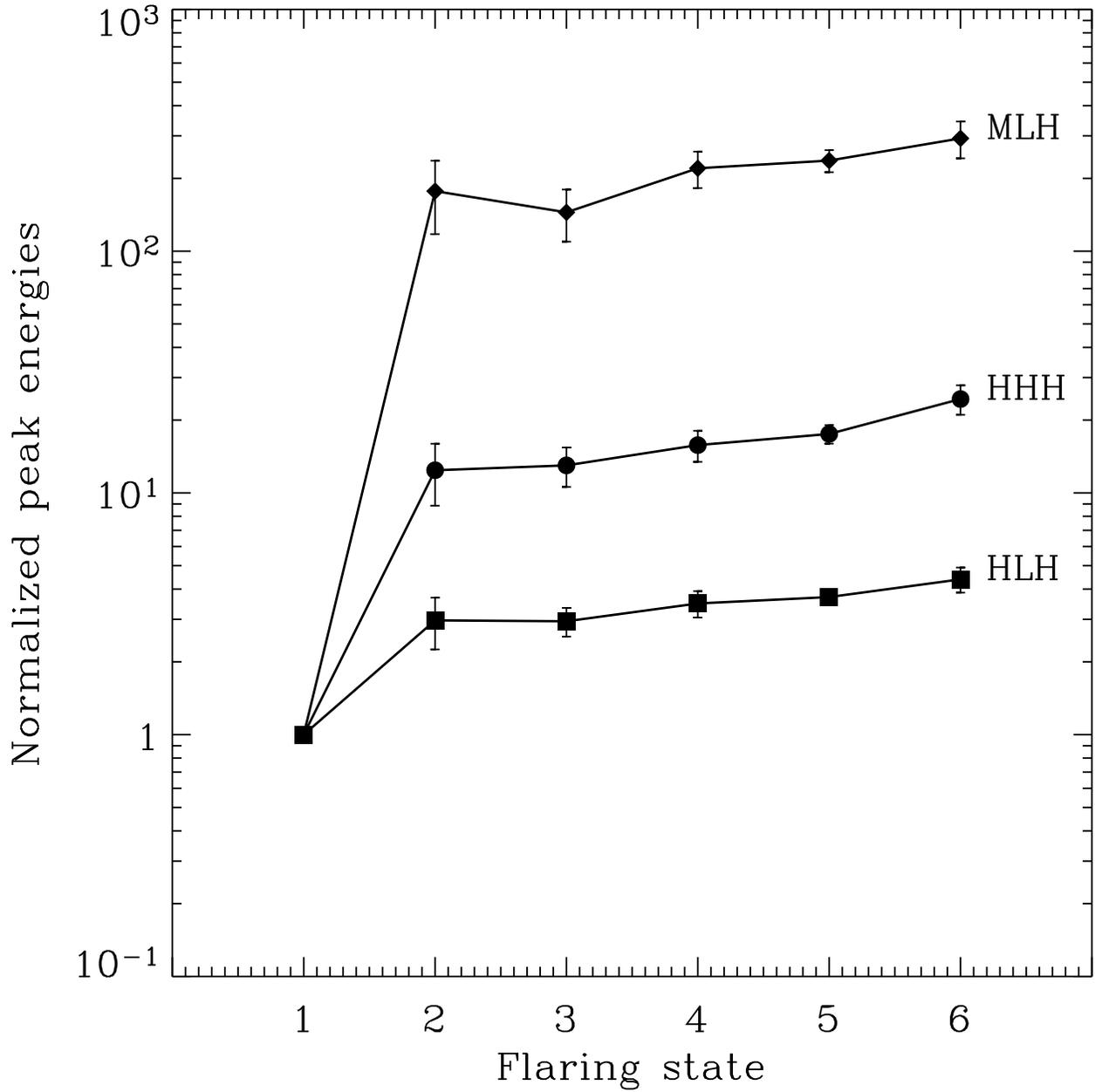}
  \end{center}
  \caption{The evolution of the peak energy of the intrinsic spectrum of Mrk~421
as a function of flaring state. All peak energies were normalized to the peak
value of the first flaring state, except for LHL, which was normalized to the
second one. The peak energies are listed in Table 6.}
  \end{figure}

\begin{figure}[t]
\begin{center}
\plotone{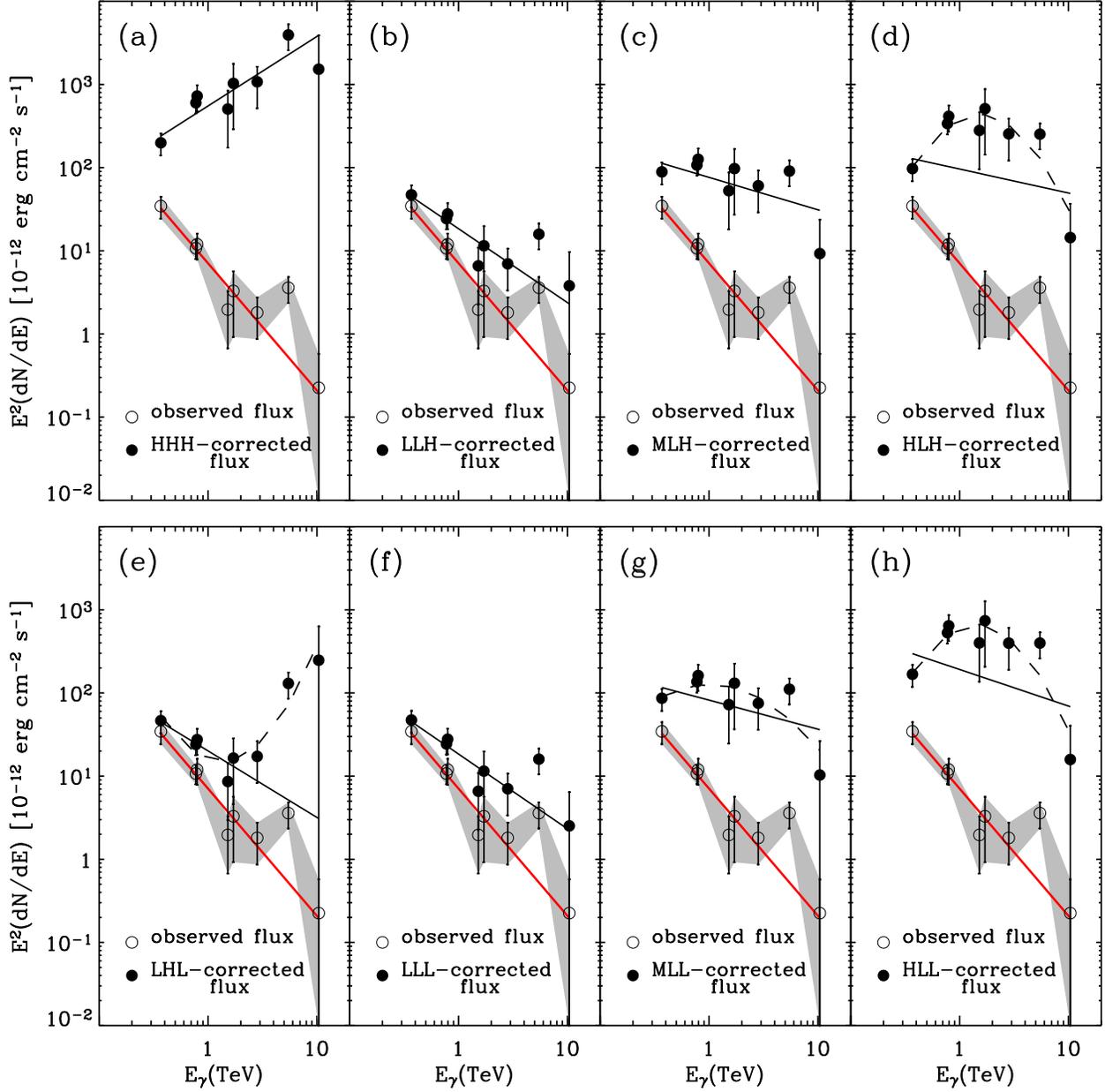}
  \end{center}
  \caption{ The observed and the absorption-corrected  spectra of the blazar
H1426+428 are plotted versus energy for the eight viable EBL realizations listed
in Table 3. The curves in the plots represent analytical approximations to the
observed and intrinsic source spectra. Observed spectra were fitted by a power
law (eq. 6 with $E_0 \rightarrow \infty$), whereas all intrinsic spectra were
fit with a parabolic function (eq. 7) and shown as solid lines. However,
intrinsic spectra corrected for the HLH, LHL, MLL, and HLL realizations of the
EBL were better fitted by an parabolic function with an exponential rise (eq.
8), and are depicted by dashed lines in the figure.} 
 \end{figure}

\end{document}